%% file: DQN_MG.tex
\documentclass[journal]{IEEEtran}
\usepackage{cite}

\ifCLASSINFOpdf
  \usepackage[pdftex]{graphicx}
  \graphicspath{{../figure/}{../jpeg/}}
  \DeclareGraphicsExtensions{.pdf,.jpeg,.png}
\else
  \usepackage[dvips]{graphicx}
  \graphicspath{{../eps/}}
  \DeclareGraphicsExtensions{.eps}
\fi

\usepackage{amsmath}

\ifCLASSOPTIONcompsoc
  \usepackage[caption=false,font=normalsize,labelfont=sf,textfont=sf]{subfig}
\else
  \usepackage[caption=false,font=footnotesize]{subfig}
\fi

\usepackage[table]{xcolor}
\usepackage{mdwmath}
\usepackage{mdwtab}
\usepackage{amssymb}
\DeclareMathOperator*{\argmax}{argmax}
\usepackage{enumerate}
\usepackage{bm}
\usepackage{multirow}
\usepackage{booktabs}
\usepackage{algorithm}
\usepackage{algorithmicx}
\usepackage{algpseudocode}
\usepackage{amssymb}
\usepackage{romannum}

\usepackage{graphics} 
\usepackage{epsfig} 
\usepackage{mathptmx} 
\usepackage{times} 
\usepackage{amsmath} 
\usepackage{amssymb}  
\usepackage{amsfonts}
\usepackage{float}
\usepackage{cite}
\usepackage{algorithm}
\usepackage[latin9]{inputenc}
\usepackage{booktabs}
\usepackage{graphicx}
\usepackage{algorithm}
\usepackage{algpseudocode}
\usepackage{comment}
\usepackage{booktabs}

\usepackage{nomencl}
\makenomenclature
\usepackage{etoolbox}
\renewcommand\nomgroup[1]{%
  \item[\bfseries
  \ifstrequal{#1}{P}{Physics Constants}{%
  \ifstrequal{#1}{N}{Number Sets}{%
  \ifstrequal{#1}{O}{Other Symbols}{}}}%
]}

\makeatletter

\newcommand{\Rmnum}[1]{\expandafter\@slowromancap\romannumeral #1@}
\makeatother

\begin{document}
\title{Double Deep Q-learning Based Real-Time Optimization Strategy for Microgrids}

\author{Hang Shuai,~\IEEEmembership{Member, ~IEEE}, Xiaomeng Ai,~\IEEEmembership{Member,~IEEE}, Jiakun Fang,~\IEEEmembership{Senior Member,~IEEE}, Wei Yao,~\IEEEmembership{Senior Member,~IEEE}, Jinyu Wen,~\IEEEmembership{Member,~IEEE}

\thanks{This work was supported by ($Corresponding$ $author:$ $Xiaomeng$ $Ai.$)}

\thanks{H. Shuai, X. Ai, J. Fang, W. Yao, and J. Y. Wen are with School of Electrical and Electronic Engineering, Huazhong University of Science and Technology, Wuhan, 430074, China, and the Department of Electrical, Computer and Biomedical Engineering, University of Rhode Island,Kingston, RI, 02881,USA. (email: xiaomengai1986@foxmail.com)}

}

\date{}

\maketitle

\maketitle

\begin{abstract}
    The uncertainties from distributed energy resources (DERs) bring significant challenges to the real-time operation of microgrids.
    In addition, due to the nonlinear constraints in the AC power flow equation and the nonlinearity of the battery storage model, etc., the optimization of the microgrid is a mixed-integer nonlinear programming (MINLP) problem.
    It is challenging to solve this kind of stochastic nonlinear optimization problem.
    To address the challenge, this paper proposes a deep reinforcement learning (DRL) based optimization strategy for the real-time operation of the microgrid.
    Specifically, we construct the detailed operation model for the microgrid and formulate the real-time optimization problem as a Markov Decision Process (MDP).
    Then, a double deep Q network (DDQN) based architecture is designed to solve the MINLP problem.
    The proposed approach can learn a near-optimal strategy only  from the historical data.
    The effectiveness of the proposed algorithm is validated by the simulations on a 10-bus microgrid system and a modified IEEE 69-bus microgrid system.
    The numerical simulation results demonstrate that the proposed approach outperforms several existing methods.
\end{abstract}

\begin{IEEEkeywords}
Real-time optimization, microgrid, deep reinforce learning (DRL), deep Q network, nonlinearity.
\end{IEEEkeywords}

\IEEEpeerreviewmaketitle

\section*{Nomenclature}
\label{sec:nomenclature}
\input{nomen}

\section{Introduction}
\label{sec:introduction}
\input{intro}

\section{Problem Formulation}
\label{sec:formulation}
\input{formulate}

\section{Proposed Approach}
\label{sec:algorithm}
\input{algorithm}

\section{Experimental Results}
\label{sec:results}
\input{results}

\section{Conclusion and Future Work}
\label{sec:conclusion}
In this paper, we proposed a DDQN based real-time optimization strategy for the optimal operation of microgrids.
The power flow constraints and the detailed battery model were considered in this paper.
To deal with the huge action space and the complex constraints in the microgrid model, we constructed the primary actions and the secondary actions.
Then, a DDQN based DRL algorithm was proposed to solve the stochastic nonlinear optimization problem.
Finally, the effectiveness of the proposed approach was validated by numerical simulations on a 10-bus microgrid test system and a modified IEEE 69-bus microgrid test system.
From the simulation results, the proposed algorithm outperformed the traditional ADP, MPC, and myopic methods in both deterministic and stochastic case studies.
Especially, we found that the DDQN-RTO algorithm can learn from the historical data to make near-optimal decisions in the real-time optimization process.

The DDQN based optimization approach needs to discretize the continuous actions in the system which is not convenient to the application of the algorithm to much complex systems.
To overcome this deficiency, the authors will combine the proposed method with other state-of-the-art reinforcement learning approach such as Trust Region Policy Optimization (TRPO) in the future work.

\ifCLASSOPTIONcaptionsoff
  \newpage
\fi

\bibliographystyle{IEEEtran}
\bibliography{IEEEabrv,DQN_MG_Ref}

\end{document}

%% file: nomen.tex
\addcontentsline{toc}{section}{Nomenclature}
\begin{IEEEdescription}[\IEEEusemathlabelsep\IEEEsetlabelwidth{$V_1,V_2,V_3,V_4$}]
\setlength{\parskip}{1pt}

\item[$\textbf{Acronyms}$]
\item[$DE$] Diesel engine generator.
\item[$DGs$] Distributed generators.
\item[$DQN$] Deep Q network.
\item[$DRL$] Deep reinforcement learning.
\item[$MINLP$] Mixed-integer nonlinear programming.
\item[$MT$] Micro-gas turbine generator.
\item[$PV$, $WT$] Photovoltaic and wind turbine.
\item[$SOC$] State of charge.

\vspace{0.2cm}
\item[$\textbf{Parameters}$]
\item[$a_g, b_g, c_g$] Fuel cost coefficients of controllable DG $g$.
\item[$C_r$] Rated capacity of battery, $kWh$.
\item[$c_{bat}$] Per KWh degradation price for battery, $\$/kWh$.
\item[$G_{i,j}, B_{i,j}$] Real part and imaginary part of the nodal admittance matrix, respectively.
\item[$\Delta t$] Time step.
\item[$ \theta $, $ \hat{\theta}$] Weights of evaluate neural network and target neural network.

\vspace{0.2cm}
\item[$\textbf{Variables}$]
\item[$a$] Action or decision variable.
\item[$a^p, a^s$] Primary actions and secondary actions.
\item[$ L $] Discharging or charging loss of the battery.
\item[$P$, $Q$] Active ($kW$) and reactive power ($kVar$) generation of power sources, such as controllable DGs, utility grid, battery, PV panels, and WT.
\item[$p$] Electricity price, $\$/kWh$.
\item[$r$] Single time period reward, $\$$.
\item[$s$] State variable.
\item[$V$] Voltage amplitude of bus in the microgrid.
\item[$\delta$] Phase angle of bus in the microgrid.
\item[$\eta$] Discharge or charge efficiency of the battery.
\item[$\omega$] Uncertainty variables of the system.

\vspace{0.2cm}
\item[$\textbf{Functions}$]
\item[$f(\cdot)$] Transition function in MDP.
\item[$ \bm{P}_{.} ( \cdot , \cdot  ) $] State transition probability.
\item[$ Q (\cdot, \cdot) $] Action-value function.
\item[$ \bm{R}_{.} ( \cdot , \cdot  ) $] Immediate reward.

\vspace{0.2cm}
\item[$\textbf{Superscripts}$]
\item[$c,d$] Charge and discharge mode of battery.
\item[$max$,$min$] Maximum and minimum value.
\item[$on$,$off$] ON and OFF status of controllable DGs.
\item[$*$] Optimal value.

\vspace{0.2cm}
\item[$\textbf{Indices}$]
\item[$bat$] Battery.
\item[$g$] Controllable DGs, such as MT and DE.
\item[$grid$] Utility grid.
\item[$i,j$] Bus index.
\item[$l$] Power cable.
\item[$n$] Iteration index.
\item[$pv,wt$] PV panel and wind generator.
\item[$t$] Time index.
\item[$\bm{\pi}$] Operational policy mapping from the state $ s $ to the action $ a $.

\end{IEEEdescription}

%% file: intro.tex
\IEEEPARstart{M}{icrogrid} is a group of interconnected loads and distributed generators (DGs) that acts as a single controllable entity with respect to the grid \cite{DOE2011}.
Recent researches show that microgrids can increase the efficiency of the power supply, improve reliability of the power system, and enable the integration of distributed energy resources (DERs) \cite{izadian2013renewable,wang2015coordinated}.
However, the random and intermittent characteristics of DERs bring significant challenges to the optimal operation of microgrids.

To deal with the uncertainty in microgrids, there have been rich prior works \cite{Optimal6184357,8666177Huang,zhang2013robust,8057282} which focused on the energy management of microgrids, and researchers have proposed plenty of optimization approaches \cite{8666177Huang,zhang2013robust,8057282}, such as robust optimization \cite{zhang2013robust} and stochastic programming \cite{8057282}.
Specifically, Robust optimization uses prior knowledge to model uncertainty parameters in a predefined uncertainty set \cite{hedman2014application}.
Stochastic programming deals with uncertainty by sampling a set of scenarios that are typically generated by Monte Carlo technique.
Then, the stochastic problem is formulated as a deterministic problem.
In generally, theses approaches formulate the optimization of microgrid as an off-line optimization problem, which makes it hard to adopt these methods to real-time scheduling.

In recent years, real-time optimization of microgrids has attracted much attention \cite{Dynamic4538359,Hang_TSG_2019,gu2017online,8658002Kong,ma2018distributed,shi2017real}.
Many real-time optimization methods, such as dynamic programming (DP) \cite{Dynamic4538359}, approximate dynamic programming (ADP) \cite{Hang_TSG_2019}, model predictive control (MPC) \cite{gu2017online,8658002Kong}, alternating direction method of multipliers (ADMM) \cite{ma2018distributed}, and Lyapunov optimization \cite{Dynamic6595653, shi2017real}, have been applied in microgrids.
Although DP is a good optimization method, it is time-consuming when applied in complex environment.
Moreover, traditional DP is difficult to adapt to any changing probabilities or un-modeled uncertainties in the environment \cite{shi2017real}.
MPC is a widely used real-time optimization approach.
But, it needs the near future forecasting information of the system.
And the linear or mixed-integer linear optimization model is usually adopted in MPC approach, since nonlinear problems are intrinsically more difficult to solve \cite{bertsekas1997nonlinear}.
In comparison with MPC, the algorithms proposed in \cite{ma2018distributed,shi2017real} do not require any forecasting information, which decreases the influence of the forecast error on the decision making process.
In \cite{shi2017real}, Lyapunov optimization relaxes the time-coupled constraints and dynamically solves the single time-period problem only according to the current system state.
However, the quadratic power flow constraints are nonlinear equations and they are hard to be dealt with using the approach proposed in \cite{shi2017real}.
Thus, the original nonlinear optimization problem is relaxed to a convex optimization problem.
Last but not least, the historical data are not fully utilized in the above real-time optimization methods.

Generally speaking, the above real-time optimization methods applied in microgrids at least have one of the following drawbacks.
\begin{enumerate}
  \item The method requires certain forecasting information. The real-time optimization performance may deteriorate due to the forecasting error.
  \item The method needs to assume a stationary stochastic process with known distribution information for the renewable energy and/or demand side \cite{shi2017real,Rahbar6913531}. It is difficult for the method to adapt to changing stochastic processes.
  \item The historical data of the system are not fully utilized.
  \item Because it is tough to solve multi-time period nonlinear programming problems \cite{bertsekas1997nonlinear}, the method usually simplify or relax the optimization model to reduce the computational cost.
\end{enumerate}

In order to overcome the drawbacks of the traditional optimization methods, \emph{learning-based optimization approaches} have been proposed to solve linear/nonlinear optimization problems \cite{IoT8664581,shuai2018stochastic,8736401He,shuai2020real, gao2014machine,Intelligent8561208}.
Reinforcement learning (RL) is a widely used learning-based decision-making method which has been applied in many aspects like the communication resource allocation \cite{IoT8664581} and nonlinear optimization in smart grid \cite{shuai2018stochastic,8736401He,shuai2020real}.
But traditional RL algorithms generally require manually designed features.
With the breakthrough of the deep learning (DL), the deep neural network has became a powerful feature extractor and has achieved great success in speech and image recognition \cite{abdel2014convolutional,Wan8561167}, and natural language processing, etc.
To combine the advantages of the traditional RL and DL method, researchers proposed the deep reinforcement learning (DRL) approachs, such as deep Q network (DQN) \cite{mnih2013playing}, deep deterministic policy gradient (DDPG) \cite{lillicrap2015continuous}, asynchronous advantage actor-critic (A3C) \cite{mnih2016asynchronous}, soft actor critic (SAC) \cite{haarnoja2018soft}, AlphaZero \cite{silver2017mastering2}, and the effectiveness were validated in complex real-time strategy games.
Realizing the advantages of the DRL approach, some researchers made the effort to use the DRL to solve the optimal operation/control problems in power industry in the past few years \cite{8661739,wan2018model,chen2017modeling,mocanu2018line,huang2019adaptive}, including the electrical vehicle (EV) charging/discharging scheduling \cite{wan2018model}, building energy management \cite{chen2017modeling,mocanu2018line}, and power system emergency control\cite{huang2019adaptive}, etc.

Regarding microgrid optimization problems, the DRL based optimization approaches are also developed in some literature recently \cite{franccois2016deep,DDQN8742669,ji2019real}.
Francois-Lavet et al. proposed a deep Q-learning based approach for the battery and hydrogen tank storage system scheduling in a microgrid \cite{franccois2016deep}.
In \cite{DDQN8742669}, a double deep Q-learning based algorithm was proposed to optimize the operation of the battery system in a microgrid.
A deep Q-learning based approach was applied in the energy management of a microgrid in \cite{ji2019real}.
Although these methods achieved promising results, they generally formulated the problem by either linearizing the component model or neglecting some constraints in a microgrid.
To obtain applicable real-time operation decisions, formulating a high fidelity microgrid model is critical \cite{olivares2014centralized}.
Constraints like AC power flow equations and detailed device models in a microgrid need to be considered in the real-time optimization process.

Unlike the above approaches that generally ignore the operational constraints in the microgrid, in this paper, we consider all the necessary operational constraints (for example, the detailed battery operation constraints, the AC power flow equations, the ON/OFF constraints of micro-generators). To the best of our knowledge, this is the first time that these constraints have been considered in DRL based microgrid real-time optimization approach.

In this paper, we propose a DRL based real-time optimization strategy for the optimal operation of the microgrid.
More Specifically, a double deep Q-learning based real-time optimization algorithm is proposed.
The battery storage system model and all the necessary physical constraints such as power flow equations are carefully considered.
The complex constraints in the microgrid cannot be directly dealt with by the neural network, so a constraint processing mechanism is proposed in this work.
We use a deep neural network to approximate the action-value function, and the proposed DRL based optimization algorithm can learn to determine an optimal operation strategy from the historical data.
Numerical simulations demonstrate the effectiveness of the proposed algorithm.

The contributions of this paper are threefold.
\begin{enumerate}
  \item  A double deep Q network based real-time optimization (DDQN-RTO) algorithm for the microgrid is proposed. The proposed algorithm takes necessary nonlinear constraints such as the AC power flow equations into consideration.
  \item To bridge the gap between deep reinforcement learning theory and the formulated MINLP problem, a novel procedure is proposed to deal with the constraints in the microgrid model.
  \item The simulation results demonstrate that the proposed DDQN-RTO algorithm can directly learn an effective real-time operation strategy by utilizing the historical data without depending on forecasting information.
\end{enumerate}

The remainder of this paper is organized as follows.
The mathematical model of a microgrid is formulated in Section \ref{sec:formulation}.
Then a deep reinforcement learning based real-time optimization strategy for microgrid is proposed in Section \ref{sec:algorithm}.
In Section \ref{sec:results}, numerical simulations are presented to demonstrate the effectiveness of the proposed approach.
Conclusions are summarized in section \ref{sec:conclusion}.

%% file: formulate.tex
\subsection{Optimal Operation of the Microgrid}
\begin{figure}[!hbt]\centering
\includegraphics[width=3in]{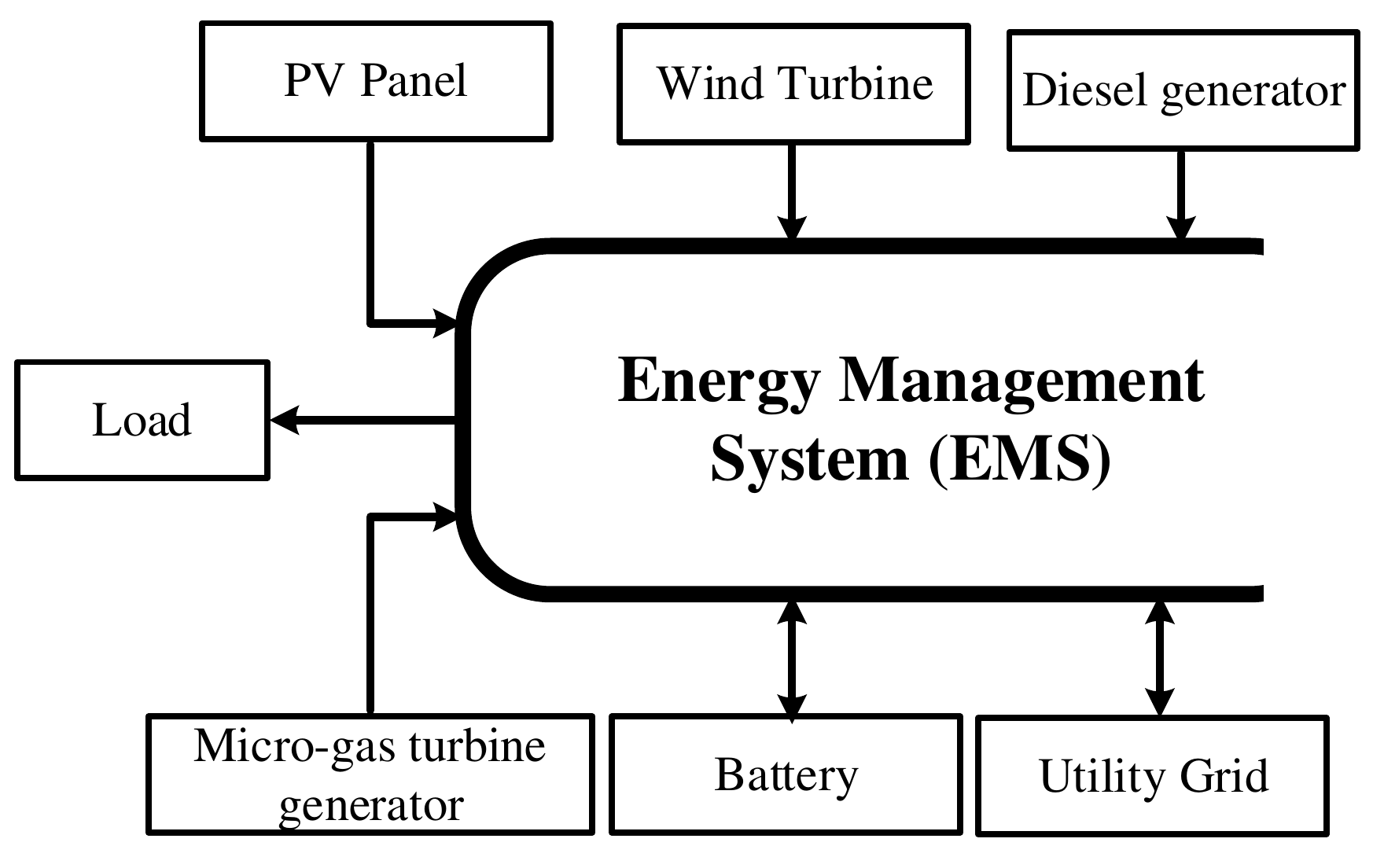}
\caption{The components of a microgrid.} \label{fig:Microgrid_Sch}
\end{figure}

In this paper, the optimal operation model of the microgrid is investigated.
A typical microgrid is shown in Fig. \ref{fig:Microgrid_Sch} which contains the diesel engine generator (DE), micro-gas turbine generator (MT), wind turbines, rooftop PV panels, and a battery storage.
In this work, the microgrid can exchange power with the utility grid.

\subsection{Learning Problem Formulation}
We formulate the real-time optimization of the microgrid as a Markov Decision Processes (MDP) with discrete time step  $ t = \{1,2,\dots,T\}$. The objective is to find the optimal schedules to minimize the total costs.

The MDP formulation has five elements, including the system state $ \bm{S} $, the set of actions $\bm{A}$, the system state transition probability $ \bm{P}_{.} ( \cdot , \cdot  ) $, the immediate reward $ \bm{R}_{.} ( \cdot , \cdot  ) $, and the discount factor $ \gamma $. The details of each element are presented as follow.

\subsubsection{State}
The system state at time $ t $ is shown in Eq. (\ref{eq:state}).

\begin{equation}
\label{eq:state}
s_t=( s_{g,{t-1}}, P_{g,{t-1}}, P_{pv,t}, P_{wt,t}, D_t, Q_t, p_t, SOC_t )
\end{equation}
where $ s_{g,{t-1}} $ and $ P_{g,{t-1}} $ represent the unit commitment and the generation dispatch of controllable DGs at previous time step;
Unit commitment state, $s_{g,{t-1}}$, is a binary variable which represents the ON/OFF status of the controllable generator $g$ at time $t-1$.
$ P_{pv,t} $ and $ P_{wt,t} $ denote the generation dispatch of PV panel and WT at current time;
$ D_t $ and $ Q_t $ represent the active and reactive power demand of the system;
$ p_t $ is the electricity price when exchanging electricity with the main grid;
$ SOC_t $ is the state of charge (SOC) of the battery.

In this work, we also consider the various constraints of the microgrid, including output limits of controllable DGs, ramping rates of controllable DGs, minimum ON/OFF time limits of controllable DGs, power flow constraints, voltage constraints, power cable capacity constraints, and battery operational constraints.
The controllable DGs in this work include the DE and MT.

\noindent \emph{Output limits of controllable DGs:} 
\begin{equation}\label{GrindEQ4}
\left\{
    \begin{array}{lcl}
    P_{g}^{min} \le P_{g,t} \le P_{g}^{max} \\
    Q_{g}^{min} \le Q_{g,t} \le Q_{g}^{max}
    \end{array}
    \right.
    \qquad \forall g , \forall t
\end{equation}

\noindent \emph{Ramping rates of controllable DGs:}
\begin{equation}\label{GrindEQ5}
\left\{
    \begin{array}{lcl}
    P_{g,t} - P_{g,{t-1}} \le & s_{g,{t}} R_{up,g} + P_{g}^{min}(s_{g,t}-s_{g,{t-1}}) \\
    &+P_{g}^{max}(1-s_{g,t}) \\
    P_{g,{t-1}} - P_{g,t} \le & s_{g,t-1} R_{dn,g} + P_{g}^{min}(s_{g,{t-1}}-s_{g,t}) \\
    &+P_{g}^{max}(1-s_{g,{t-1}})
    \end{array}
    \right.
    \quad \forall g , \forall t
\end{equation}
where $ R_{up,g} $ and $ R_{dn,g} $ are the ramping up and ramping down rates of controllable DG $g$, respectively.

\noindent \emph{Minimum ON/OFF time limits of controllable DGs:} 
\begin{equation}\label{GrindEQ6}
\left\{
    \begin{array}{lcl}
    (s_{g,{t-1}}-s_{g,t})(S_{g,{t-1}}^{on} -T_{g}^{on}) \ge 0 \\
    (s_{g,t}-s_{g,{t-1}})(S_{g,{t-1}}^{off} -T_{g}^{off}) \ge 0
    \end{array}
    \right.
    \quad \forall g , \forall t
\end{equation}
where $ S_{g,{t-1}}^{on} $ and $ S_{g,{t-1}}^{off} $ are the ON and OFF time duration of the DG until time $t-1$, respectively; $T_{g}^{on}$ and $T_{g}^{off}$ are the minimum ON and OFF duration of the DG, respectively.

\noindent \emph{Power flow constraints:}
\small
\begin{equation}\label{GrindEQ3}
\begin{aligned}
    V_{i,t} \sum_{j=1}^{N_{bus}} V_{j,t} ( G_{ij} cos \delta_{ij,t} + B_{ij} sin \delta_{ij,t} ) = \sum_{s=1}^{N_s} I_{i,s} P_{s,t} - D_{i,t}  \quad \forall t \\
    V_{i,t} \sum_{j=1}^{N_{bus}} V_{j,t} ( G_{ij} sin \delta_{ij,t} - B_{ij} cos \delta_{ij,t} ) = \sum_{s=1}^{N_s} I_{i,s} Q_{s,t} - Q_{i,t} \quad \forall t \\
\end{aligned}
\end{equation}
\normalsize
where $ s $ denotes the different power sources in the microgrid, including WT, PV, MT, DE, battery, and utility grid.
$N_{bus}$ is the total bus number of the system.
$N_{s}$ is the total bus number that {\color{red}is} connected to power sources.
$I_{i,s}$ is the element of the bus - generator correlation matrix.
$D_{i,t}$ and $Q_{i,t}$ are the active ($kW$) and reactive ($kVar$) load of the bus $i$ at time $t$, respectively.
$G_{i,j}$ and $B_{i,j}$ are the real part and imaginary part of the nodal admittance matrix, respectively.

\noindent \emph{Voltage amplitude and phase angle constraints:}
\begin{equation}\label{GrindEQ7}
\left\{
    \begin{array}{lcl}
        V_{i}^{min} \le V_{i,t} \le V_{i}^{max} \\
        \delta_{i}^{min} \le \delta_{i,t} \le \delta_{i}^{max}
        \end{array}
    \right.
    \qquad \forall i , \forall t
\end{equation}

\noindent \emph{Power cable capacity constraints:}
\begin{equation}\label{GrindEQ8}
    P_{l,t} \le P_{l}^{max} \qquad \forall l , \forall t
\end{equation}
where $P_{l,t}$ and $P_{l}^{max}$ represent the power transmission and the maximum power transmission of cable $l$, respectively.

\noindent \emph{Battery constraints:}

Battery SOC constraints are expressed as Eq. (\ref{GrindEQ18}).

\begin{equation}\label{GrindEQ18}
    SOC^{min} \le SOC_t \le SOC^{max}
\end{equation}

In this work, the battery discharging/charging constraints are also considered. The discharging loss $ L_{bat,t}^d $ and charging loss $ L_{bat,t}^c $ are shown in Eq. (\ref{GrindEQ9}) and (\ref{GrindEQ10}) \cite {Nguyen2016}.

\small
\begin{equation}\label{GrindEQ9}
\begin{aligned}
    L_{bat,t}^d & = \frac {10^3 \left( R_{in} + \frac{K_b}{SOC_t} \right)} {V_r^2}  P_{bat,t}^2 \\
    & + \frac {10^3 C_r . K_b \left( 1-SOC_t \right)}{SOC_t.V_r^2} P_{bat,t}
\end{aligned}
\end{equation}
\normalsize

\small
\begin{equation}\label{GrindEQ10}
\begin{aligned}
    L_{bat,t}^c &= \frac {10^3 \left( R_{in} + \frac{K_b}{1.1-SOC_t} \right)} {V_r^2} P_{bat,t}^2 \\
    &- \frac {10^3 C_r . K_b \left( 1-SOC_t \right)}{SOC_t.V_r^2} P_{bat,t}
\end{aligned}
\end{equation}
\normalsize
where $R_{in}$ is the internal resistance of the battery;
$V_r$ and $C_r$ are the rated voltage and rated capacity of the battery respectively;
$K_b$ is the polarisation constant of battery;
where $P_{bat,t}$ is the battery discharging/charging power, which is positive during discharging and negative during charging.
Therefore, the discharging and charging efficiencies can be expressed as Eq. (\ref{GrindEQ11}) and (\ref{GrindEQ12}).

\begin{equation}\label{GrindEQ11}
    \eta_t^{d}=\frac {P_{bat,t}}{ P_{bat,t} + L_{bat,t}^d}
\end{equation}
\begin{equation}\label{GrindEQ12}
    \eta_t^{c}=1 + \frac {L_{bat,t}^c}{P_{bat,t}}
\end{equation}

According to the laboratory and field tests \cite{SAKTI2017279}, the battery discharging/charging power limits are not only related to the maximum discharging power $ P^{d,max} $ and minimum charging power $ P^{c,min} $ but also determined by the SOC. The discharging/charging power limits are expressed as Eq. (\ref{GrindEQ13}) and (\ref{GrindEQ14}).

\begin{equation}\label{GrindEQ13}
P_{bat,t} \le \min \left\lbrace  P_{bat,t}^{d,max} (SOC_t,\eta_t^{d,min}), P^{d,max} \right\rbrace
\end{equation}

\begin{equation}\label{GrindEQ14}
P_{bat,t} \geq \max \left\lbrace P_{bat,t}^{c,min} (SOC_t,\eta_t^{c,min}), P^{c,min}  \right\rbrace
\end{equation}
where $ P_{bat,t}^{d,max} (.) $ and $ P_{bat,t}^{c,min} (.) $ can be calculated as Eq. (\ref{GrindEQ15}) and (\ref{GrindEQ16}).

\begin{equation}\label{GrindEQ15}
    P_{bat,t}^{d,max} (SOC_t,\eta_t^{d,min})= \frac{V_r^2 SOC_t (\frac{1}{\eta_t^{d,min}}-1) - 10^3 C_r K_b (1-SOC_t)}{10^3(R_{in} \cdot SOC_t + K_b)}
\end{equation}
\normalsize
\begin{equation}\label{GrindEQ16}
    P_{bat,t}^{c,min} (SOC_t,\eta_t^{c,min})= \frac{10^3 C_r K_b (1-SOC_t) - SOC_t V_r^2 (1- \eta_t^{c,min})}{10^3 \cdot SOC_t (R_{in} + \frac{K_b}{1.1-SOC_t})}
\end{equation}
\normalsize

\subsubsection{Action}
The action is defined as follows:
\begin{equation}
\label{eq:action}
a_t = ( s_{g,t}, P_{g,t}, Q_{g,t}, P_{grid,t}, Q_{grid,t}, P_{bat,t}, Q_{bat,t}, V_{i,t}, \delta_{i,t} )
\end{equation}
where $ s_{g,t} $ represents the unit commitment of controllable DGs; $P_{g,t}$ and $Q_{g,t}$ are the active and reactive power generation of controllable DGs; $P_{grid,t}$ and $Q_{grid,t}$ are the active and reactive power exchange between the microgrid and the external grid; $ P_{bat,t} $ is the battery charging/discharging power; $ Q_{bat,t} $ is the reactive power generation of the battery; $V_{i,t}$ and $\delta_{i,t}$ are the magnitude and the phase angle of the $i$th bus, respectively.
The subscript $t$ represents the current time period.

\subsubsection{State transition}
The state transition from the current state $s_t$ to the next state $s_{t+1}$ is defined as

\begin{equation}
\label{eq:state_transition}
s_{t+1}=f(s_t,a_t, \omega_t)
\end{equation}
where the state transition is determined by the action $a_t$ and the uncertainty $\omega_{t}$. Specifically, the state transition for the battery SOC is determined by Eq. (\ref{eq:SOC_transition}).

\begin{equation}\label{eq:SOC_transition}
SOC_{t}=\left\{
    \begin{array}{lr}
    SOC_{t-1}-\frac{P_{bat,t} + L_{bat,t}^d}{C_r} \Delta t, & \  {P_{bat,t} > 0}    \\
    SOC_{t-1}+\frac{-P_{bat,t } - L_{bat,t}^c}{C_r} \Delta t, &  {P_{bat,t} < 0}   \\
    \end{array}
\right.
\end{equation}
where $\Delta t$ is the time step.
In this work, we set $\Delta t$ equals to 1 hour.

For the state transition of the unit commitment of the controllable DGs, it is directly controlled by the action $ a_t $. For $ P_{pv,t}, P_{wt,t}, D_t, Q_t, p_t $, their state transitions are subject to randomness because the generation dispatch of WT and PV panel, the active and reactive load, and the electricity price are all unknown in advance.
For the transition of the generation dispatch of DG, it is determined by the optimization process.

\subsubsection{Reward}
The reward at each time step is calculated as Eq. (\ref{eq:reward}).

\begin{equation}\label{eq:reward}
\begin{aligned}
    r_t =  -s_{g,t} ( a_g (P_{g,t} \Delta t)^2 + b_g P_{g,t} \Delta t + c_g )\\
    - C_{sup,g} - p_t P_{grid,t} \Delta t - C_{bat,t}
\end{aligned}
\end{equation}
This reward consists of four parts: the first part $ s_{g,t} ( a_g P_{g,t}^2 + b_g P_{g,t} + c_g ) \Delta t $ shows the fuel cost of controllable DGs, and it is a quadratic function of its generation dispatch; the second part $ C_{sup,g} $ represents the startup cost of controllable DGs; the third part $ p_t P_{grid,t} \Delta t $ denotes the electricity cost when purchasing electricity from the main grid; the fourth part $ C_{bat,t} $ is the battery degradation cost that is calculated according to \cite {Nguyen2016} and is shown in Eq. (\ref{GrindEQ19}).

\begin{equation}\label{GrindEQ19}
C_{bat,t}=\left\{
    \begin{array}{lr}
    c_{bat} (P_{bat,t} + L_{bat,t}^d) \Delta t,& \ {P_{bat,t} > 0}  \\
    c_{bat} L_{bat,t}^c \Delta t, & \ {P_{bat,t} < 0} \\
    \end{array}
    \right.
\end{equation}

\subsubsection{Action-value function}
The quality of taking an action $ a $ at a state $ s $ is measured by the expected total sum of future rewards as Eq. (\ref{eq:action_value_function}).

\begin{equation}
\label{eq:action_value_function}
 Q_{\bm{\pi}} (s, a) = \mathbb{E}_{\bm{\pi}} \left. \left[ \sum_{k=t}^K \gamma^{k-t} \cdot r_{k} \right| s_t = s, a_t = a \right]
\end{equation}
where $ Q_{\bm{\pi}} (s, a) $ denotes the action-value function; $\bm{\pi}$ represents the operational policy mapping from the state $ s $ to the action $ a $; $ \gamma $ is the discount factor that is applied to balance the importance between the immediate reward and future rewards; $ K $ is the optimization horizon.

The objective of the microgrid optimal operation is to achieve an optimal policy $\bm{\pi}^* $ such that the action-value function can be maximized as Eq. (\ref{eq:max_action_value_function}).

\begin{equation}
\label{eq:max_action_value_function}
Q^{\ast} (s, a) = \max_{\bm{\pi}} Q_{\bm{\pi}} (s, a)
\end{equation}
In the above equation, $ Q^{\ast} (s, a) $ represents the optimal action-value function.

%% file: algorithm.tex
This work focuses on the real-time optimization of microgrids.
Due to the nonlinear AC power flow constraints (5), the nonlinear battery model (8) - (16), and the binary variables in actions (17), etc., the optimization model is a nonlinear optimization problem.
In addition, the renewable energy introduces uncertainty into the optimization problem.
These make it challenging for the traditional model-based approaches to get the optimal operation policy.
A RL-based approach is proposed in this paper to solve this problem.
The proposed approach can directly learn an optimal policy from the historical data.

Specifically, the RL approach updates the action-value function based on the Bellman equation as:
\begin{equation}
\label{eq:Bellman Eq}
Q_{n+1} (s, a) = \mathbb{E}\Big\{r_t + \gamma \max_{a_{t+1}}Q_n(s_{t+1},a_{t+1})|s_t = s, a_t = a\Big\}
\end{equation}
where $n$ is the iteration index.
Theoretically, $Q(s,a)$ will converge to the optimal function when the number of iteration $n$ approaches infinite value\cite{sutton1998reinforcement}.
Then, the optimal operation policy can be determined by:
\begin{equation}
\label{eq:OptimalAction}
a^{*} = \arg \max_{a}Q^{*}(s,a)
\end{equation}

A deep neural network is used to approximate the action-value function. With the combination of deep neural network and the RL, a deep reinforcement learning based approach is proposed in this paper to solve the real-time optimization problem for microgrid.

\subsection{Architecture of the Proposed DRL based Microgrid Optimization}
\begin{figure}[!hbt]\centering
\includegraphics[width=3.2in]{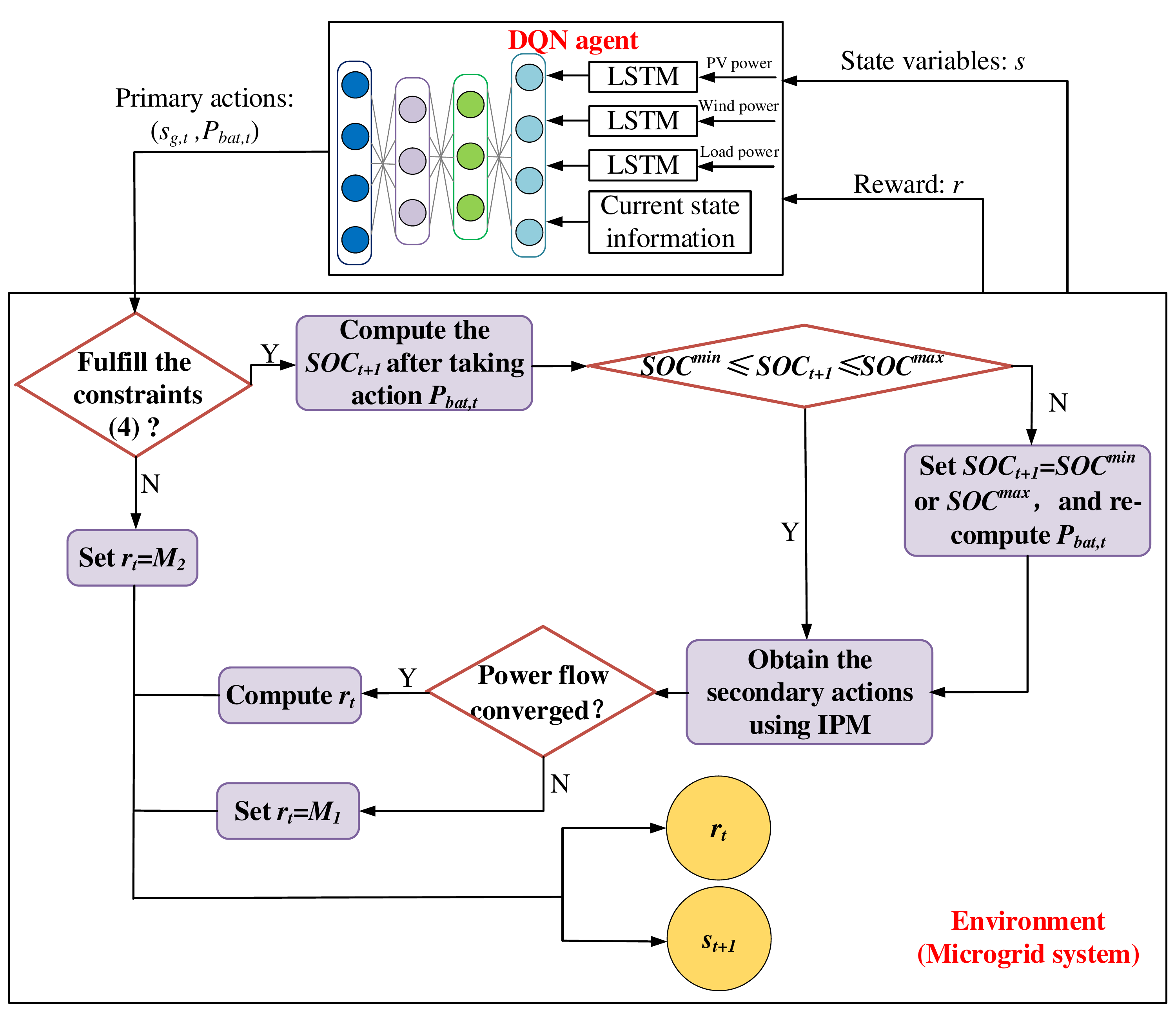}
\caption{The proposed architecture of the proposed algorithm.} \label{fig:DRL}
\end{figure}

The architecture of the proposed DRL based microgrid optimization approach is shown in Fig. \ref{fig:DRL}.
The DQN based agent determines the optimal actions of the microgrid according to the state information of the system.
Then, the actions are executed in the environment, i.e., the microgrid system, and the next system state and the reward of the action are obtained.
In this procedure, all the necessary constraints shown in Section II are considered.
The reward signal is fed into the agent to guide the agent to make a better decision in the next iteration.
The details of the architecture are unfolded in the following parts.

Neural network (NN) can uniformly approximate any continuous functions.
The state information of the microgrid, $s$, is fed to the input layer of the DQN, then by computing forward through the layers of the NN we can obtain the output value $Q{(s,a)}$.
\begin{equation} \centering
\label{eq:DQN feedforward}
\begin{aligned}
v_{t}^{in}&=g(W_1 * s_t + b_1) \\
v_{t}^{h1}&=g(W_2 * v_{t}^{in} + b_2) \\
&\vdots \quad\quad\quad\quad \\
v_{t}^{hm}&=g(W_m * v_{t}^{hm-1} + b_m) \\
Q(s,a) &= g(W_{m+1} * v_{t}^{hm} + b_{m+1})
\end{aligned}
\end{equation}
where $v$ is the value of the hidden unit, $g(\cdot)$ is the activation function, $W$ is the matrix of weights, $b$ is the vector of biases, and $Q{(s,a)}$ is the output of the DQN which denotes the action-value for all feasible actions when the microgrid under state $s$.

The number of the output neurons of the DQN is determined by the action space.
From Eq. (\ref{eq:action}), the decisions $P_{g,t}$, $Q_{g,t}$, $P_{grid,t}$, $Q_{grid,t}$ $V_{i,t}$, and $\delta_{i,t}$ are all continuous variables and the dimensions of the variables increase quickly with the number of the generators and buses in the microgrid.
If all these actions are included in the output layer of the DQN, the NN will become very large.
Besides, the actions made by the DRL algorithm must fulfill the constraints Eq.(\ref{GrindEQ4}) - Eq.(\ref{GrindEQ16}).
These constraints are pretty hard to deal with, which will bring significant challenges to the design of the DQN and the solving of Eq.(\ref{eq:max_action_value_function}).

To decrease the dimension of the output neurons of the DQN, we divide the original actions into the primary actions and the secondary actions.
The primary actions include the ON/OFF actions of the DGs and the charge/discharge power of the battery, as given by:
\begin{equation}
\label{eq:primary actions}
a_t^{p} = \big\{s_{g,t}, P_{bat,t}\big\}
\end{equation}

The secondary actions include the generation dispatch of DGs, the power exchange between the microgrid and the external grid, the reactive power generation of the battery bus, and the magnitude/phase angle of the buses:
\begin{equation}
\label{eq:secondary actions}
a_t^{s} = \big\{P_{g,t}, Q_{g,t}, P_{grid,t}, Q_{grid,t}, Q_{bat,t}, V_{i,t}, \delta_{i,t}\big\}
\end{equation}
The primary actions are determined by solving Eq.(\ref{eq:max_action_value_function}).
While the secondary actions are obtained by solving the optimal power flow problem by considering the constraints Eq.(\ref{GrindEQ4}) - Eq.(\ref{GrindEQ16}) using the interior point method (IPM) after the primary actions have been determined.
The details of the decision making process are shown in Fig. \ref{fig:DRL}.
According to the state information of the microgrid, the primary actions are determined by the DQN agent using the following equation:
\begin{equation}
\label{eq:primary actions making}
a_t^{p} = \arg \max{Q(s,a_t^{p})}
\end{equation}
Then, the DQN based real-time optimization algorithm needs to ensure that the primary actions fulfill the minimum ON/OFF time limits shown in Eq. (\ref{GrindEQ6}).
When the constraints are not fulfilled, the reward is set to be a small value $M_2$.
After receiving a small reward, the agent learns that the action should obey Eq. (\ref{GrindEQ6}).
Thus, in the following training iterations, the agent will try to make sure the DG's ON/OFF time duration larger than the limitation to get a bigger reward.
Otherwise, using the charge/discharge power given by the primary actions, the SOC of the battery at the next time period can be calculated by Eq. (\ref{eq:SOC_transition}).
If the SOC of the battery exceeds the upper or lower boundary, we set the SOC equals to $SOC^{max}$ or $SOC^{min}$ and recompute the output power of the battery.
Then, the active/reactive power output of all controllable and uncontrollable DGs, and the voltage of all buses are determined by solving the optimal power flow (OPF) subproblem.
In this work, the OPF is calculated by IPM.
If the power flow converged, we could obtain the optimal power output of DGs and the external grid, so the reward can be calculated by Eq. (\ref{eq:reward}).
Otherwise, the reward is set to be a small number $M_1$.

Fig. \ref{fig:DRL} shows that the proposed architecture greatly decreased the action space.
More importantly, a large part of the operational constraints of the microgrid is handled by IPM.
These make it possible to use DQN agent to solve this problem.

\subsection{Training Process of the DDQN-RTO Algorithm}
The DQN agent needs to be trained off-line first.
The training process of the proposed algorithm is shown in Algorithm 1.
To overcome unstable learning process, experience replay \cite{mnih2013playing} technique is adopted in Algorithm 1.
Experience replay mechanism works by storing historical transitions $(s_t, a_t, r_t, s_{t+1})$ in the replay memory and randomly sampling a mini-batch data from the stored transitions to train the NN.
Randomizing the samples breaks the temporal correlations between the system state transitions and therefore reduces the variance of the learning process.
Besides, using the same neural network both to select and to evaluate an action will lead to overestimations \cite{van2016deep}.
The target neural network is introduced in Algorithm 1 to reduce overestimations.
The target Q network has the same architecture with the evaluate Q network.

The first step is to randomly initialize the parameters of the neural networks.
Then, in the outer loop the DQN is updated for $N$ epoches.
Each epoch includes $T$ training steps.
In every time step, the agent selects a action based on $\varepsilon$-greedy policy:
\begin{equation}
\label{eq:greedy policy}
a_t^p = \left\{
    \begin{array}{lr}
    \arg\max_{a_t^p \in A} {Q(s_t, a_t^p; \theta)},& \ $randn(1)$ > \varepsilon  \\
    select \ a \ random \ action, & \ $randn(1)$ < \varepsilon \\
    \end{array}
    \right.
\end{equation}
where $\varepsilon$ is gradually decreasing with the training step, as shown below
\begin{equation}
\label{eq:vare}
\varepsilon = \max (\varepsilon - \Delta \varepsilon, \varepsilon^{min})
\end{equation}
where $\varepsilon^{min}$ is the minimum value.
According to the selected primary action, the reward and the next system state can be calculated by the procedure shown in Fig. \ref{fig:DRL}.
Then the transition $(s_t, a_t, r_t, s_{t+1})$ is stored in the experience buffer $ \mathcal{D} $.
In the next, a minibatch of the history transitions are selected and used to train the DQN agent.
By minimizing the loss function $L(\cdot)$, the parameters of the evaluate Q network are updated as:
\begin{equation}
\label{eq: parametyer uupdate}
\theta_{t+1} = \theta_t - \beta \nabla_{\theta_t} L(\theta_t)
\end{equation}
where $\nabla_{\theta_t}$ is the gradient of the loss function, $\beta$ is the learning rate of the algorithm.
The loss function is the error between the target value and the evaluate value which can be defined as:
\begin{equation}
\label{eq: loss function}
L(\theta) = \sum_{j=1}^{\#\mathcal{F}}\mathbb{E} \left[ \left( y_j - Q\left( s_j, a_j^p;  \theta \right) \right)^2 \right]
\end{equation}
In Eq. (\ref{eq: loss function}), the target value is calculated as:
\begin{equation}
\label{eq: target action-value function}
y_j = r_j + \gamma \arg \max_{a^p} \hat Q(s_{j+1},a^p;\hat \theta)
\end{equation}
Finally, the parameters of the target Q network are updated.
To further stable the learning process, we adopted the soft update strategy to update the target network:
\begin{equation}
\label{eq:soft update}
\hat \theta = (1-\tau) \hat \theta + \tau \theta
\end{equation}
where $\hat \theta$ and $\theta$ are the parameters of the target Q network and evaluate Q network, respectively; $\tau$ is the updating rate.
It is worth to note that the target Q network is also updated in every step.

\begin{algorithm}[!t]
\caption{Training of Double Deep Q Network with experience replay} \label{alg:DQN}
\begin{algorithmic}[1]
\State Initialization: set the replay memory $ \mathcal{D} $; initialize action-value function $Q$ and target action-value function $\hat{Q}$ with random weights $ \theta $ and $ \hat{\theta} $, respectively.

	\For{Epoch=1:N}
		\State Generate a training scenario and initialize state $ s_1 $.
		\For{Time step t=1:T}
			\State \parbox[t]{\dimexpr\linewidth-\algorithmicindent}{ Select primary action $ a_t^p $ based on $ \varepsilon$-greedy policy.\strut}
			\State \parbox[t]{\dimexpr\linewidth-\algorithmicindent}{Execute action $a_t^p$ in emulator. Then, observe \\reward $ r_t $ and process to the new state $ s_{t+1} $ using\\ the method shown in Fig. \ref{fig:DRL}.\strut}
			\State \parbox[t]{\dimexpr\linewidth-\algorithmicindent}{Store transition $ \left( s_t, a_t^p, r_t, s_{t+1} \right) $ in $ \mathcal{D} $.\strut}
			\State \parbox[t]{\dimexpr\linewidth-\algorithmicindent}{Sample random minibatch of transitions \\ $ \mathcal{F} = \left\lbrace \left( s_j, a_j^p, r_j, s_{j+1} \right) \right\rbrace_{j=1}^{\#\mathcal{F}} $ from $ \mathcal{D} $.\strut}
			\State \parbox[t]{\dimexpr\linewidth-\algorithmicindent}{ {\small$ y_j \longleftarrow  r_j + \gamma \argmax_{a^p} \hat{Q}\left( s_{j+1}, a^p; \hat{\theta} \right) $.}\strut}
			\State  \parbox[t]{\dimexpr\linewidth-\algorithmicindent}{Update parameters $ \theta $ by minimizing loss \\ function $ \sum_{j=1}^{\#\mathcal{F}}\mathbb{E} \left[ \left( y_j - Q\left( s_j, a_j^p;  \theta \right) \right)^2 \right] $.\strut}	
	\State \parbox[t]{\dimexpr\linewidth-\algorithmicindent}{Update the parameters of the target Q network \\ using Eq. (\ref{eq:soft update})}				
		\EndFor
	\EndFor
\end{algorithmic}
\end{algorithm}

\subsection{DDQN based Real-time Optimization for the Microgrid}
After the DQN agent has been well trained by Algorithm 1, it will be directly utilized in the real-time optimization process of the microgrid.
The details of the real-time optimization process are shown in Algorithm 2.
The parameters of the DDQN won't be updated during the process.
For the current time $t$, the DQN agent obtains the system state firstly.
Then the $Q$ values are calculated for all feasible primary actions under the system state $s_t$.
The optimal primary actions will be selected and the secondary actions are obtained by solving the OPF problem.
By executing the real-time operational decisions, the system steps into the next time period.
In the next time period, the decisions can be obtained by repeating the above procedure.

\begin{algorithm}[!t]
\caption{Real-time optimization of the microgrid using the well-trained DQN agent} \label{alg:DQN}
\begin{algorithmic}[1]
\State Load the well-trained DQN's parmeters $\theta$.
	\For{Time step $t=1:T$}
			\State \parbox[t]{\dimexpr\linewidth-\algorithmicindent}{ Obtain current state information of the microgrid.\strut}
			\State \parbox[t]{\dimexpr\linewidth-\algorithmicindent}{DQN calculates the action-value $Q(s_t, a_t^p; \theta)$.\strut}
			\State \parbox[t]{\dimexpr\linewidth-\algorithmicindent}{Obtain the primary actions by $a_t^p = \arg\max_{a_t^p \in A} {Q(s_t, a_t^p; \theta)}$.\strut}
			\State \parbox[t]{\dimexpr\linewidth-\algorithmicindent}{Calculate the secondary actions by solving OPF problem using IPM.\strut}
			\State \parbox[t]{\dimexpr\linewidth-\algorithmicindent}{ {Output the optimal actions.}\strut}				
	\EndFor
\end{algorithmic}
\end{algorithm}

%% file: results.tex
The performance of the proposed DDQN-RTO algorithm was tested through simulations on a 10-bus microgrid system and a modified 69-bus microgrid system.
To validate the effectiveness of the proposed algorithm, we firstly designed a deterministic case and compared the DDQN-RTO algorithm with look-up table based approximate dynamic programming (ADP) algorithm \cite{shuai2018stochastic}, myopic policy \cite{powell2007approximate} and particle swarm optimization (PSO) algorithm, etc.
Then, the real-time optimization performance of the proposed algorithm was tested in the stochastic case study.
Finally, in order to demonstrate the algorithm can learn an optimal decision policy from historical data, we explored the optimization performance of the DDQN-RTO algorithm using the historical data from actual power grid.

\begin{figure}[!hbt]\centering
\includegraphics[width=3.5in]{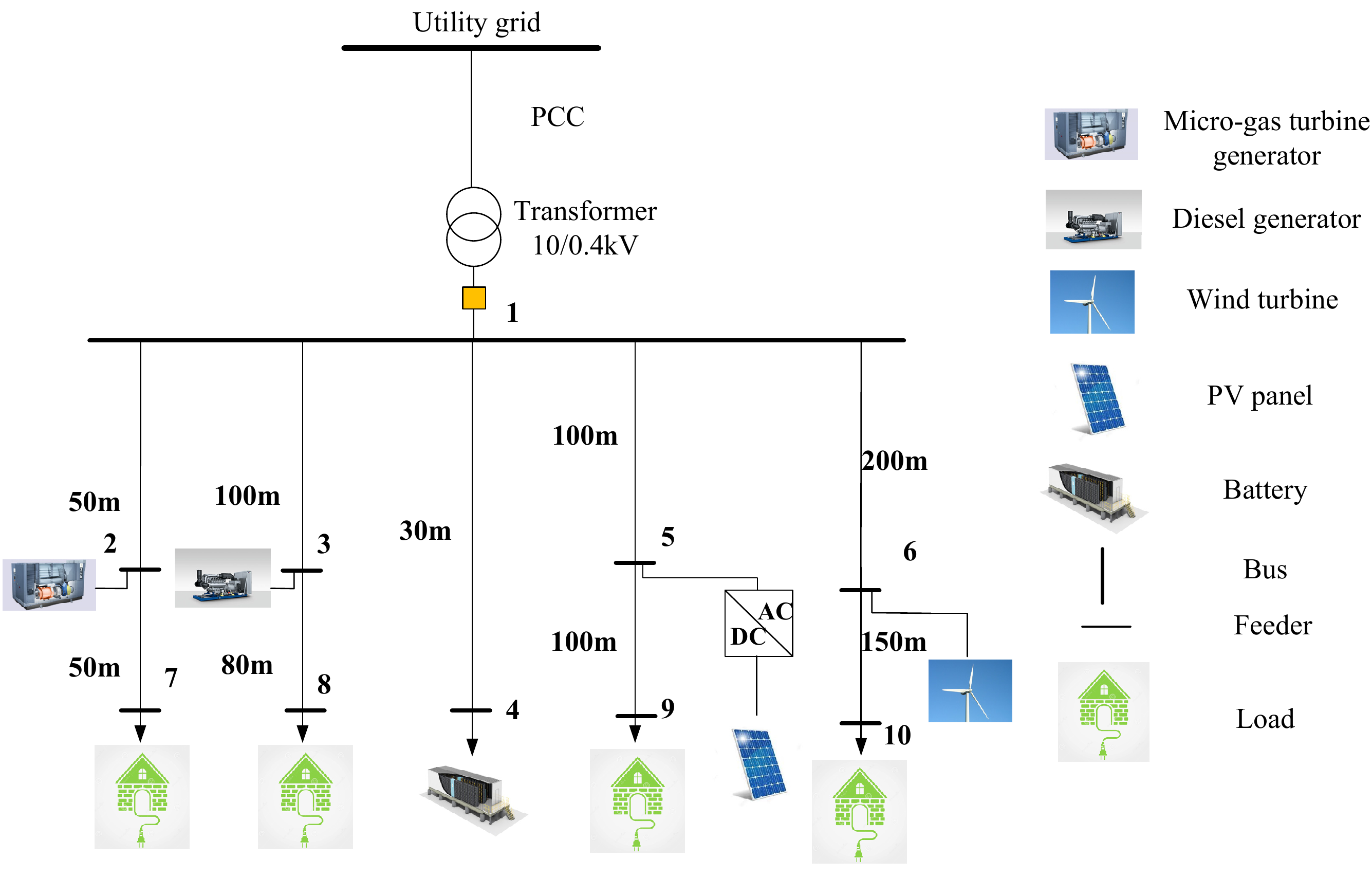}
\caption{The test microgrid system.} \label{fig:microgrid}
\end{figure}

The test microgrid system \cite{shuai2018stochastic} is shown in Fig. \ref{fig:microgrid}.
The microgrid includes a micro-gas turbine generator, a diesel generator, wind turbines, PV panels, and a battery storage system.
The maximum/minimum power output of  micro-gas turbine and the diesel generator is 30 kW and 10 kW, respectively.
The minimum start-up and shut-down time of all controllable DGs are 1 hour.
The start-up cost of the micro-gas turbine generator and diesel generator are \$ 2 and \$ 3, respectively.
The fuel cost coefficients of the controllable DGs are shown in Tab. \ref{fuelcost}.
\begin{table}\centering
\footnotesize
\caption{The fuel cost coefficients of MT and DE}\label{fuelcost}
\begin{tabular}
{|c|c|c|c|} \hline
Controllable DGs &$a$ (\$/$kWh$) &$b$ (\$/$kWh$) &$c$ (\$) \\ \hline	
MT & 0.00051	 & 0.0397    &  0.4 \\ \hline
DE & 0.00104	 & 0.0304    &  1.3 \\ \hline
\end{tabular}
\end{table}
The minimum and maximum energy can be stored in the battery are 18kWh and 60kWh, respectively.
The maximum charge/discharge power is 12 kW.
The operation cost coefficient of the battery is 0.059 \$/kWh.
Other necessary parameters of the battery can be found in \cite{shuai2018stochastic}.
The power exchange limitation between the microgrid and the external grid is 50kW.
The reactance and resistance parameters of all cables are $X = 0.1 \Omega/km$, $R = 0.64 \Omega/km$.
We implemented the proposed DDQN-RTO method in Python and the \emph{PYPOWER 5.1.4} package was adopted as the optimal power flow solver.
All case studies are conducted on a 64-bit windows based computer with 16 GB RAM and Intel Core i7 processor clocking at 3.41 GHz.

\subsection{Case Study I: Deterministic Optimization}
\begin{figure}[!hbt]\centering
\includegraphics[width=3.2in]{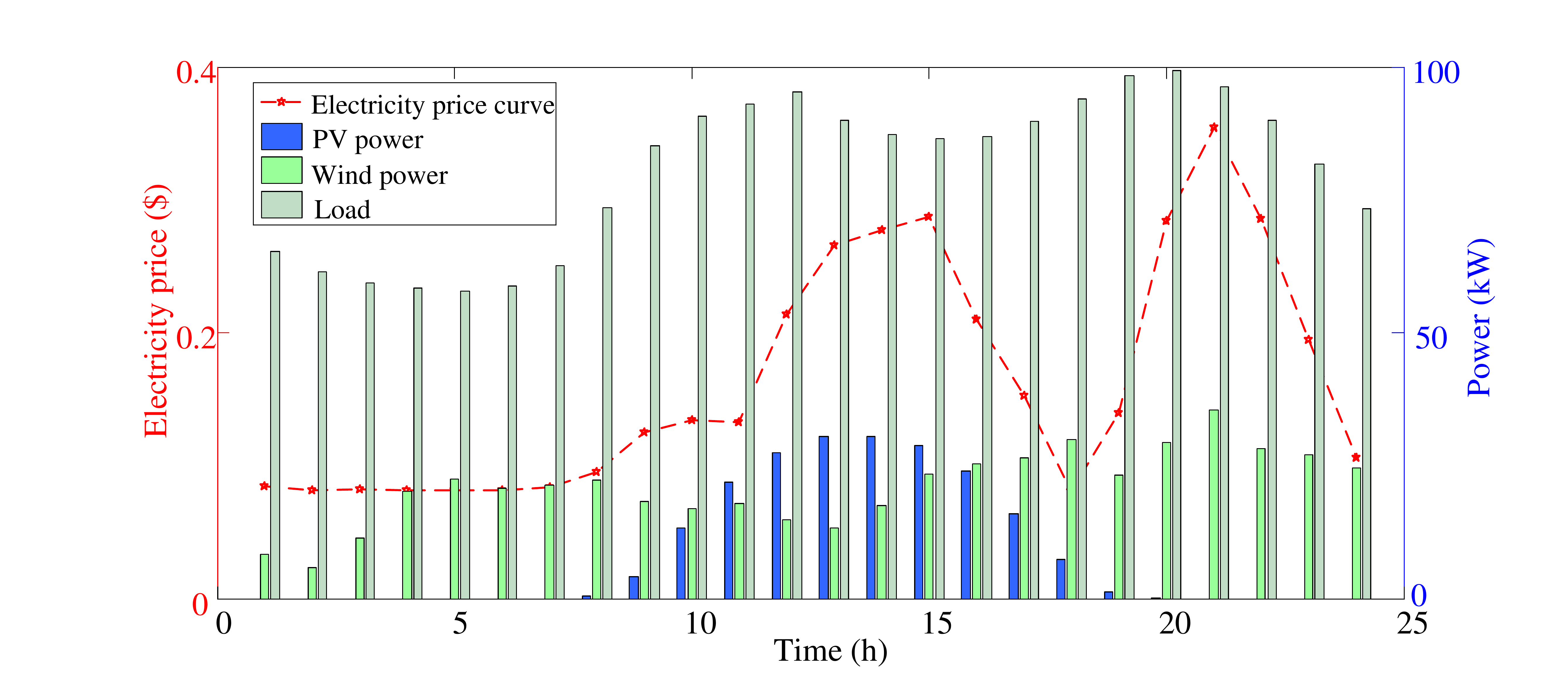}
\caption{The forecast information of the microgrid.} \label{fig:LoadWTPV}
\end{figure}

The forecasted power generation of wind turbines and PV panels are shown in Fig. \ref{fig:LoadWTPV}.
We use Algorithm 1 to train the DQN agent.
Since there is not any uncertainty exits in the deterministic case, the training scenarios are all the same in each epoch.
Part of the system states which include $( s_{g,{t-1}}, P_{pv,t}, P_{wt,t}, D_t, p_t, SOC_t )$ are directly input to the DDQN, thus the input layer has 7 neurons.
The number of layers and the number of neurons in each layer was carefully designed to obtain a good optimization performance.
Finally, the neuron network has four layers of hidden neurons, each layer having 50, 100, 100, 50 neurons, respectively.
All the neurons located in input and hidden layers are with ReLU as the activation function.
We discretize the charge/discharge power of the battery into 9 levels (-12 kW, -9 kW, -6 kW, -3 kW, 0 kW, 3 kW, 6 kW, 9 kW, 12 kW).
The output layer of the DDQN has 36 neurons and each neuron represents the Q-value of a feasible action.

The learning rate is set to $\beta = 0.01$, the discount factor to $\gamma = 0.99$, the reply memory size to $D = 10000$, the batch size to $B = 32$, $\Delta \varepsilon = 5 * 10^{-5}$, and $\varepsilon_0 = 1$.
The model is trained for 1500 episodes, where an episode is composed by 24 time steps.
The parameters of the evaluate and target Q network are updated in every step.
The Adam optimizer is adopted in the training process to minimize the loss function.

\begin{figure}[!hbt]\centering
\includegraphics[width=3.0in]{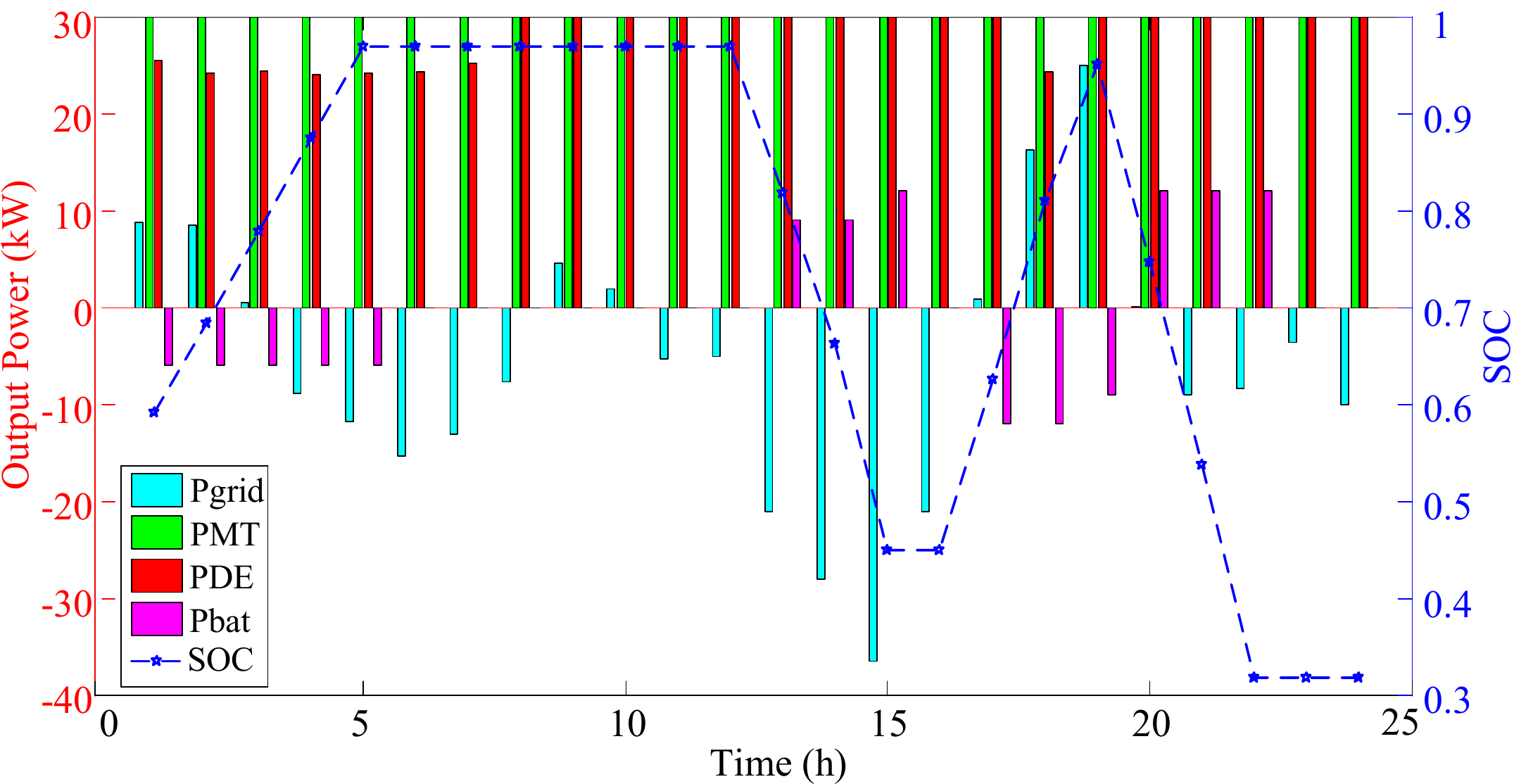}
\caption{The optimized power output of the power sources using the DDQN-RTO algorithm.} \label{fig:ResultDeterministic2}
\end{figure}

The simulation results are shown in Fig. \ref{fig:ResultDeterministic2}.
From the figure, it can be found that the load electricity is mainly provided by the MT and DE.
The microgrid needs to buy electricity from the external grid in the evening during which the demand is high, but renewable energy is relatively low.
The battery stores energy during the midnight hours and discharges in the evening peak hours.
To validate the optimization performance of the proposed DDQN-RTO algorithm, we compared the method with PSO, ADP \cite{shuai2018stochastic}, and the DP algorithm.
The total operation cost of the microgrid obtained by different methods is shown in Tab. \ref{Cost}.
In this work, the optimization result of the DP method is used as ground truth.
From the results, the DDQN-RTO algorithm performs better than myopic policy, PSO algorithm and ADP algorithm.
Thus, the effectiveness of the proposed DDQN-RTO algorithm is demonstrated in the deterministic case.
\begin{table}\centering
\footnotesize
\caption{\label{Cost}The operation cost of the microgrid}
\begin{tabular}{lcl}
\toprule
Algorithm &Operation Cost (\$) &Optimality Gap \\ 	
\midrule
DDQN-RTO & 90.25	 & 0.85\% \\
ADP & 90.26	 & 0.86\% \\
PSO & 95.68	 & 6.92\% \\
Myopic &102.68 & 14.74\% \\
DP & 89.49	 & 0\% \\
\bottomrule
\end{tabular}
\end{table}

\subsection{Case Study II: Real-Time Optimization Considering Uncertainty}
Since the load and wind/solar power exit forecasting errors, real-time optimization is needed to deal with the uncertainty.
To facilitate the DQN agent to learn the uncertainties brought by wind/solar power, electricity price, and electricity load, a sequence of the most recent forecast information including $(P_{pv,t+1}, P_{pv,t+2}, \cdots ,P_{pv,t+H})$, $(P_{wt,t+1}, P_{wt,t+2}, \cdots , P_{wt,t+H})$,  $( \\ p_{t+1}, p_{t+2}, \cdots , p_{t+H})$, and $(D_{t+1}, D_{t+2}, \cdots , D_{t+H})$, are stacked with the state vector shown in Eq. (1) and set as the input of the DDQN at time $t$.
In this case, the $H$ is set to be 4.
So the input layer of the DDQN consists of 23 neurons.
The structure of the hidden layer of the network is the same as the NN in the case study I.
The proposed DQN agent is first trained to learn the optimal operation strategy under uncertainties.
Then, the well-trained algorithm is used in the real-time optimization process and compared with several other methods.

1) $The \; Off-Line \; Training \; Process \; of \; the \; DDQN-RTO \; Algorithm$

\begin{figure}[!hbt]\centering
\includegraphics[width=3.2in]{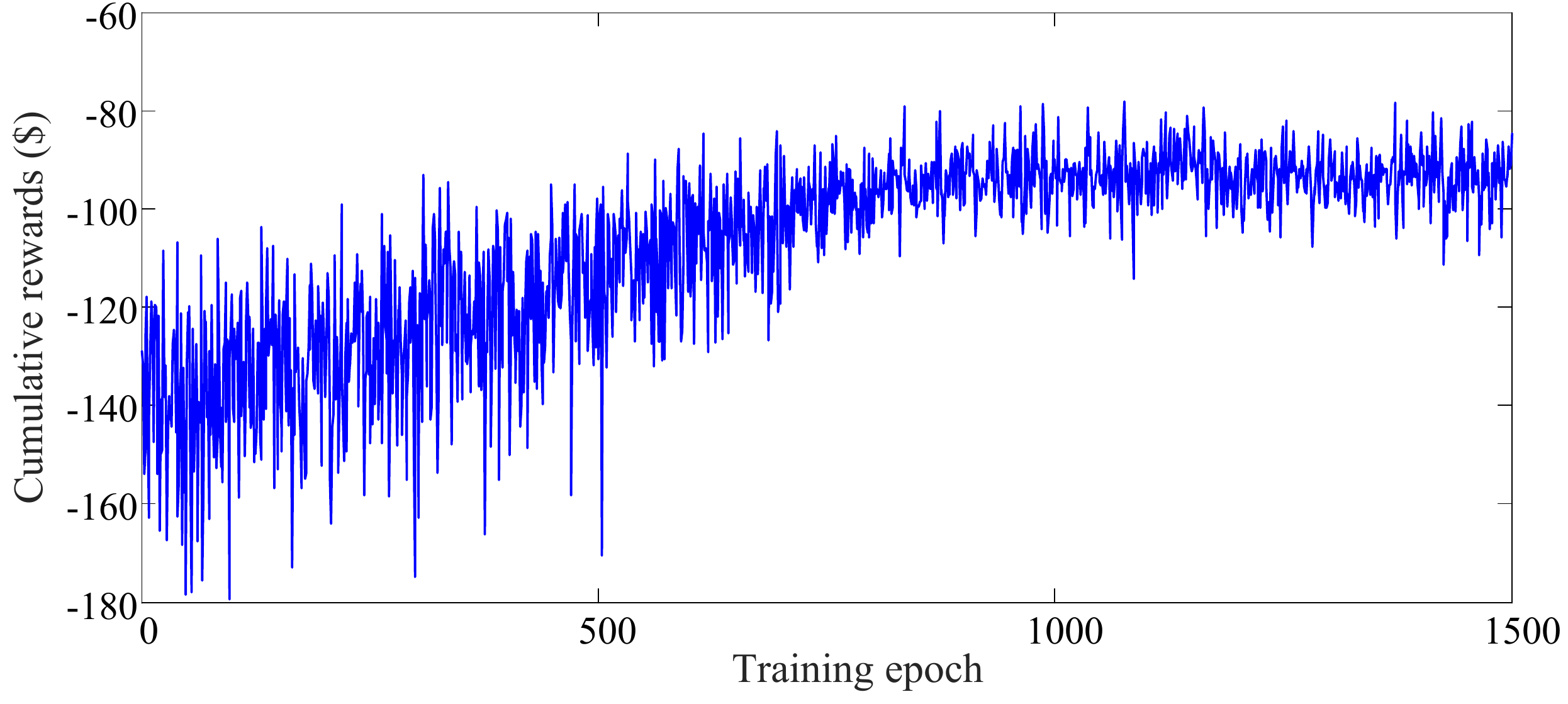}
\caption{The convergence process of the cumulative rewards.} \label{fig:Convergence1}
\end{figure}

\begin{figure}[!hbt]\centering
\includegraphics[width=3.2in]{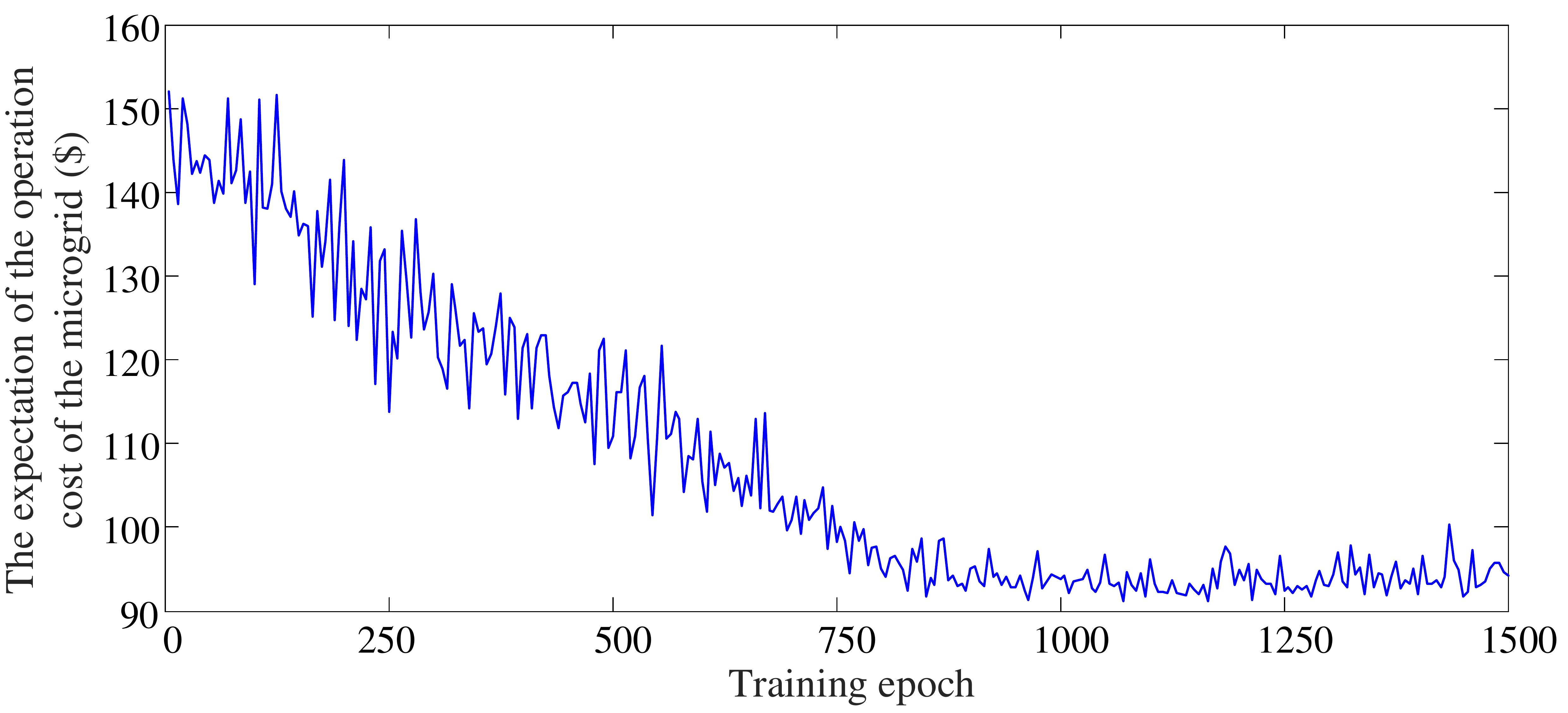}
\caption{The expected operation cost of the microgrid during the training process.} \label{fig:Convergence2}
\end{figure}

The training procedure is shown in Algorithm 1.
To train the agent, we generated 1500 set of training scenarios according to the day-ahead forecast curves and the day-ahead forecast error distribution information of wind and PV power generation, load power, and electricity price.
The scenarios are generated using the Monte Carlo simulation method.
Then, the intra-day forecasting errors will be added to the above generated scenarios to take intra-day uncertainty into consideration.
In this work, we assume the day-ahead and intra-day forecasting errors of renewable energy, electricity price, and load obey the Normal Distribution as shown in Tab. \ref{Uncertainty} \cite{8252927,7029108,7089325, 5406166}.
The proposed algorithm is trained in 1500 epochs.
In each epoch, we randomly select a training scenario from the generated scenarios in step 3 of Algorithm 1, and the cumulative rewards of the scenario are calculated.
Besides, we select 50 test scenarios to calculate the expected operation cost of the microgrid using the updated DQN agent every 5 training epochs.
The cumulative rewards over 1500 training epochs are shown in Fig. \ref{fig:Convergence1}.
The expected operation cost of the microgrid over the training epochs is shown in Fig. \ref{fig:Convergence2}.
From the results,  the DDQN-RTO algorithm converged after 1500 training epochs.

\begin{table}\centering
\footnotesize
\caption{\label{Uncertainty}The forecasting error distribution}
\begin{tabular}{lcl}
\toprule
Forecasting error distribution &Day-ahead  &Intra-day \\ 	
\midrule
Wind power & $\mathcal {N}(0,0.1^2)$	 & $\mathcal {N}(0,0.05^2)$ \\
PV power & $\mathcal {N}(0,0.1^2)$	 & $\mathcal {N}(0,0.05^2)$ \\
Load & $\mathcal {N}(0,0.05^2)$	 & $\mathcal {N}(0,0.02^2)$ \\
Electricity price & $\mathcal {N}(0,0.05^2)$	 & $\mathcal {N}(0,0.03^2)$\\
\bottomrule
\end{tabular}
\end{table}


2) $The \; Real-Time \; Optimization \; Performance \; of \; the \; DDQN-RTO \; Algorithm$

\begin{figure}[!hbt]\centering
\includegraphics[width=2.8in]{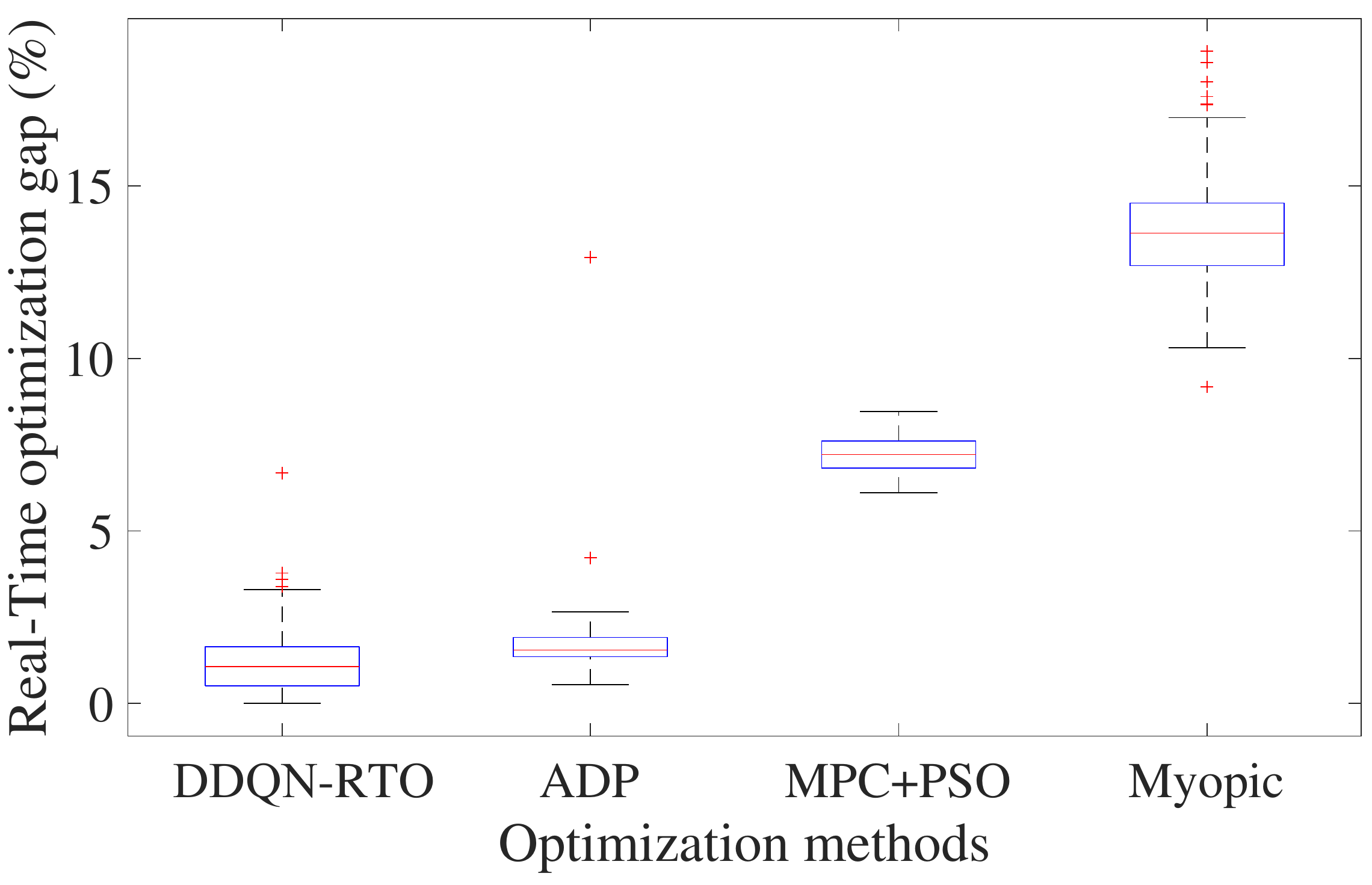}
\caption{The real-time optimization gaps of the DDQN-RTO, ADP, MPC+PSO, and myopic methods.} \label{fig:Boxplot}
\end{figure}

After the DQN agent was well-trained day-ahead, we can use it to obtain the real-time operational decisions according to the actual state of the system.
To validate the real-time optimization performance of the proposed algorithm, we randomly selected 200 test scenarios.
Similar to the training scenarios, the Monte Carlo simulation is adopted to generate the test scenarios.
For every scenario, the online optimization process is shown in Algorithm 2.
We also compared the performance of the DDQN-RTO algorithm with ADP, MPC+PSO algorithm, and myopic policy.
It is worth to note that the real-time optimization model is a MINLP problem.
Thus the traditional MPC method is hard to be directly used to solve the problem.
So the MPC embedded with PSO method which indicated by MPC+PSO is adopted to solve the nonlinear problem.
For a single test scenario, we use the result optimized by DP as the baseline.
The simulation results are shown in Fig. \ref{fig:Boxplot}.
The average optimality gap of the DDQN-RTO is 1.23\%.
While the average optimality gap of the ADP, MPC+PSO algorithm, and myopic policy are 1.80\%, 7.22\%, and 13.75\%, respectively.
We can find that the average optimization performance of the DDQN-RTO algorithm outperforms the other three algorithms.
Compared with the value table based function approximation method adopted by the ADP algorithm, the learning ability of DQN agent is more powerful.
Thus the proposed algorithm can obtain better real-time optimization performance in the stochastic environment.
MPC+PSO policy gets the operational decisions on a shorter optimization window which makes the optimization result is not global optimality.
Although the MPC+PSO method also utilized intra-day updated forecasting information, the proposed learning-based algorithm performs better.

\subsection{Case Study III: Real-Time Optimization by Learning from Historical Data}
The real-time optimization process shown in case study II needs to train the DQN agent according to the day-ahead forecasting information and the forest error distributions about the DERs, load, and electricity price.
Then, system operators use the well-trained DQN agent to optimize the operational decisions of the microgrid in real-time.
Thus, the DQN agent needs to be trained every day.
Unfortunately, this training procedure is very time-consuming.
For example, the total training process shown in Fig. \ref{fig:Convergence1} takes about 10 hours.
Besides, the forecasting error distribution information is also needed to generate training scenarios.
However, finding out the accurate forecasting error distributions of renewable energy and load is still not a easy task.
Considering the load demand and the power generation of PV are periodic, the DQN agent can be trained by the historical data and be utilized in the real-time optimization in the future.
In this section, we will train the DQN agent by using the historical renewable generation data \cite{WTPV_Data} and load power data \cite{LoadData} from the open power system data platform and further investigate the effectiveness of the proposed algorithm in the forecasting free circumstance.

\begin{figure}[!hbt]\centering
\includegraphics[width=3.2in]{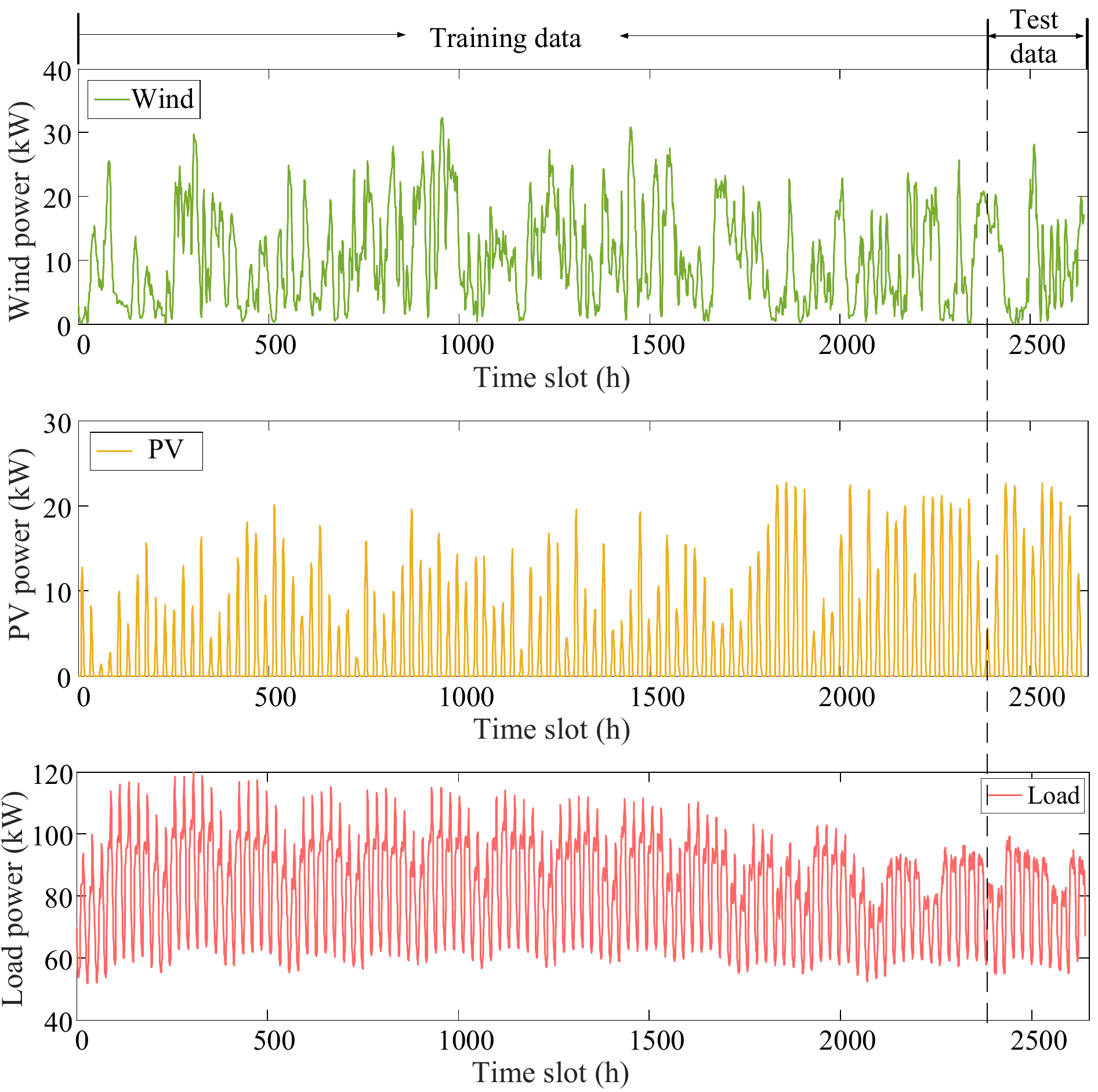}
\caption{The training and test data.} \label{fig:TrainStochasticData}
\end{figure}

The data used in this case is shown in Fig. \ref{fig:TrainStochasticData}, and the electricity price is shown in Fig. \ref{fig:LoadWTPV}.
We used 100 days of historical data to train the agent and 10 days of data to test the real-time optimization performance of the DDQN-RTO.
To learn features from historical data, the LSTM technique is used to extract features.
The inputs of the LSTM include the past 24 hours load, and power generation of WT and PV.
The outputs of the LSTM are fed to the input layer of the DDQN.
In this work, 3 LSTM networks are adopted to extract features from the historical wind/solar power, and load curves.
The number of the output of each LSTM is 64.
The output of LSTM and the current system state $( s_{g,{t-1}}, P_{pv,t}, P_{wt,t}, D_t, p_t, SOC_t )$ are input to the DDQN.
The DQN has two hidden layers, and each layer has 256 neurons.
The output layer of the DDQN also has 36 neurons.
The hyperparameters of the DDQN are the same as the above case study.
The training process is similar with the procedures given in Algorithm 1, and we randomly select a day and use the corresponding data to train the agent in each epoch.
Every 10 training epochs, we use the selected test scenarios to test the performance of the proposed algorithm.

\begin{figure}[!hbt]\centering
\includegraphics[width=3.5in]{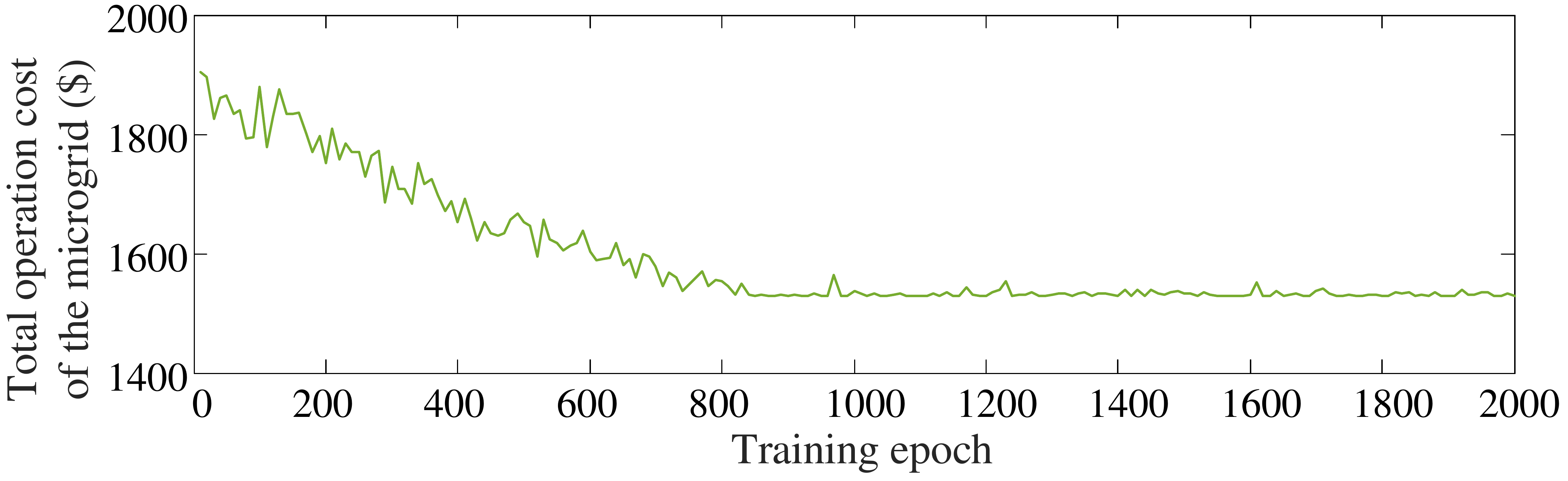}
\caption{The convergence process of the DDQN-RTO algorithm.} \label{fig:TotalCostTestConvergence}
\end{figure}

\begin{figure*}[!hbt]\centering
\includegraphics[width=6in]{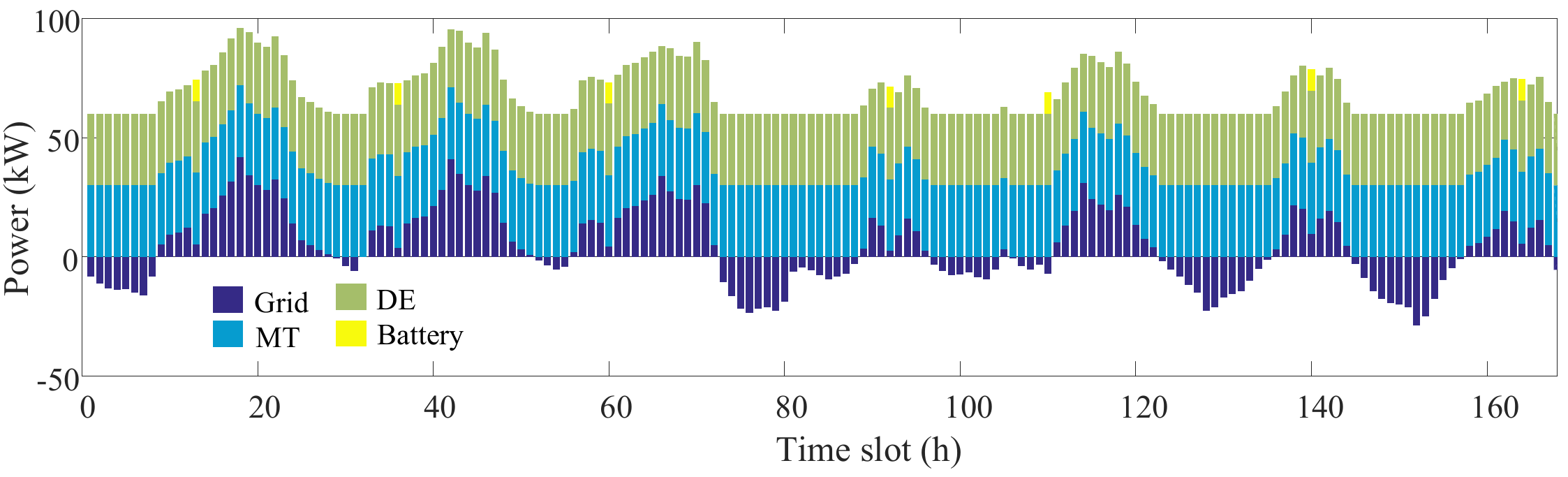}
\caption{The power output of the DGs and the utility grid.} \label{fig:PowerOutputSto}
\end{figure*}

Fig. \ref{fig:TotalCostTestConvergence} shows the convergence process of the DDQN-RTO algorithm.
It can be seen from the result that the objective function gradually decreased with the training process.
Finally, the operational cost of the microgrid converged around \$ 1524.49.
The result illustrates that the proposed DDQN-RTO algorithm can successfully learn a policy to minimize the operational cost of the microgrid.
Then, we use the well-trained DQN agent to test the real-time optimization performance of the DDQN-RTO algorithm under the test scenarios.
The generation dispatch of all DGs over 7 consecutive days are shown in Fig. \ref{fig:PowerOutputSto}.
The fuel consumption costs of MT and DE are lower than electricity price.
We can observe that the proposed DDQN-RTO algorithm has learned to provide the load by DGs first and to purchase electricity from the utility grid in the peak load hours and sell surplus power to the grid in the midnight.

\begin{figure}[!hbt]\centering
\includegraphics[width=3.2in]{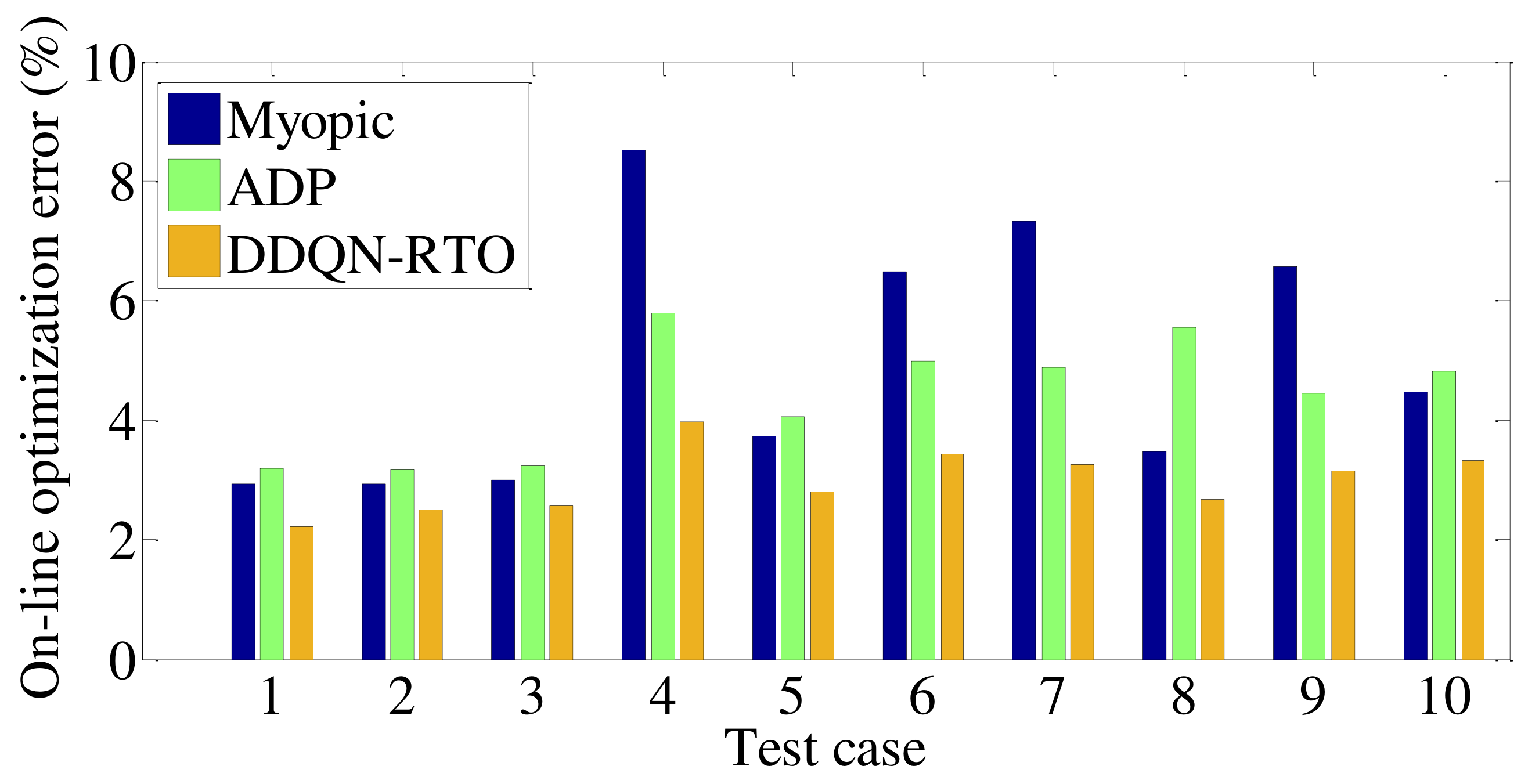}
\caption{The optimality gap of the DDQN-RTO and the ADP algorithm.} \label{fig:OnlineError_Testcase}
\end{figure}

To further validate the effectiveness of the proposed algorithm, we compared the real-time optimization performance of the DDQN-RTO algorithm with ADP algorithm and myopic policy.
For each test scenario, the optimal operation cost can be calculated by the DP algorithm.
Thus, the optimization result of DP is used as ground truth.
The operation costs optimized by different methods are shown in Table \ref{Test scenarios}.
The optimality gaps of the three algorithms are also shown in Fig. \ref{fig:OnlineError_Testcase}.
The average optimality gap of the DDQN-RTO, ADP algorithm, and myopic policy are 2.98\%, 4.41 \%, and 4.94\%, respectively.
From the results, it can be found that the DDQN-RTO algorithm performs better than the ADP algorithm and the myopic policy.

\begin{table*}\centering
\footnotesize
\caption{\label{Test scenarios}The operation cost of the microgrid under the test scenarios (\$)}
\begin{tabular}{lcllllllllll}
\toprule
Test Scenario &\#1 &\#2 &\#3 &\#4 &\#5 &\#6 &\#7 &\#8 &\#9 &\#10 \\ 	
\midrule
DDQN-RTO & 184.19	 & 195.98 & 191.83 & 110.43 & 154.02 & 128.06 & 122.33 & 165.78 & 140.36 & 131.50 \\
ADP & 185.95	 &197.27 & 193.12 & 112.36 & 155.88 &129.99 & 124.26 & 170.44 & 142.12 &133.42 \\
Myopic &185.49 & 196.81 & 192.65 & 115.25 & 155.42 & 131.84 & 127.15 & 167.08 & 145.02 & 132.96\\
DP & 180.20	 & 191.20 & 187.04 &106.21 &149.81 & 123.82 & 118.47 & 161.47 & 136.07 & 127.28 \\
\bottomrule
\end{tabular}
\end{table*}

\subsection{Case Study IV: Real-Time Optimization Performance of the DDQN-RTO Algorithm on the Modified IEEE 69-bus Microgrid System }
\begin{figure*}[!hbt]\centering
\includegraphics[width=6in]{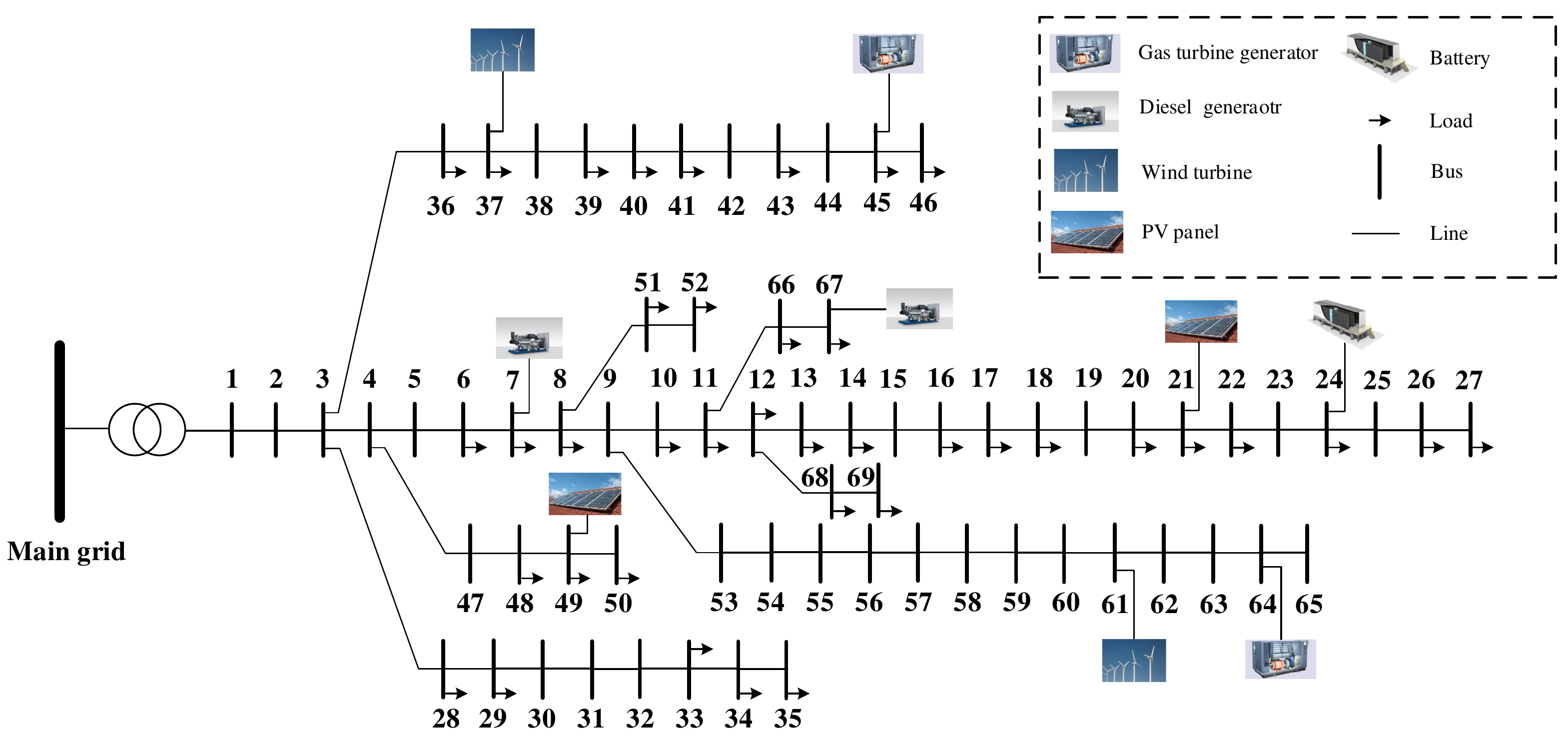}
\caption{The single-line diagram of the IEEE 69-bus microgrid system.} \label{fig:IEEE69BusSystem}
\end{figure*}

To further demonstrate the proposed algorithm can learn to operate the system from historical data, we tested the performance of the DDQN-RTO algorithm on a relatively larger microgrid system, as shown in Fig. \ref{fig:IEEE69BusSystem}.
The test system is modified based on the standard IEEE 69-bus distribution system \cite{IEEE69busData}.
In the test system, there includes a battery storage system and 8 micro-generators which including distributed PV panels, wind turbines, gas turbine generators, and diesel generators.
The locations of the micro-generators have been shown in Fig. \ref{fig:IEEE69BusSystem}.
The maximum power output of the diesel generator and micro-gas turbine generator are 450 kW, and the minimum value are 150 kW.
The capacity of the wind farm on bus 37 and PV panels on bus 21 are 525 kW and 465 kW, respectively.
The total capacity of the wind power and the solar power in the system are 1050 kW and 930 kW, respectively.
The maximum charge/discharge power of the battery in Fig. \ref{fig:IEEE69BusSystem} is 360 kW, and the capacity is 1800 kWh.
The power transmission limitation between the microgrid and the main grid is 1500 kW.
The maximum load power is 3000 kW.
The grid network parameters are modified accordingly in this work based on the parameters provided by \cite{IEEE69busData}.

We discretize the charge/discharge power of the battery into 9 levels (-360kW, -270kW, -180kW, -90kW, 0kW, 90kW, 180kW, 270kW, 360kW).
The commitment number of the controllable micro-generators is 16.
So, the size of the decision space is 144.
Then, the output layer of the DDQN used in this simulation includes 144 neurons.
The neural network architecture and the hyperparameter of the LSTM and DDQN are the same with Case study III.
The training data and test date used in this case are similar with the data used in Case study III, which are generated by scaling up the curves in Fig. \ref{fig:TrainStochasticData} according to the capacity of the renewable energy and the maximum load power of the system.

\begin{figure}[!hbt]\centering
\includegraphics[width=3.5in]{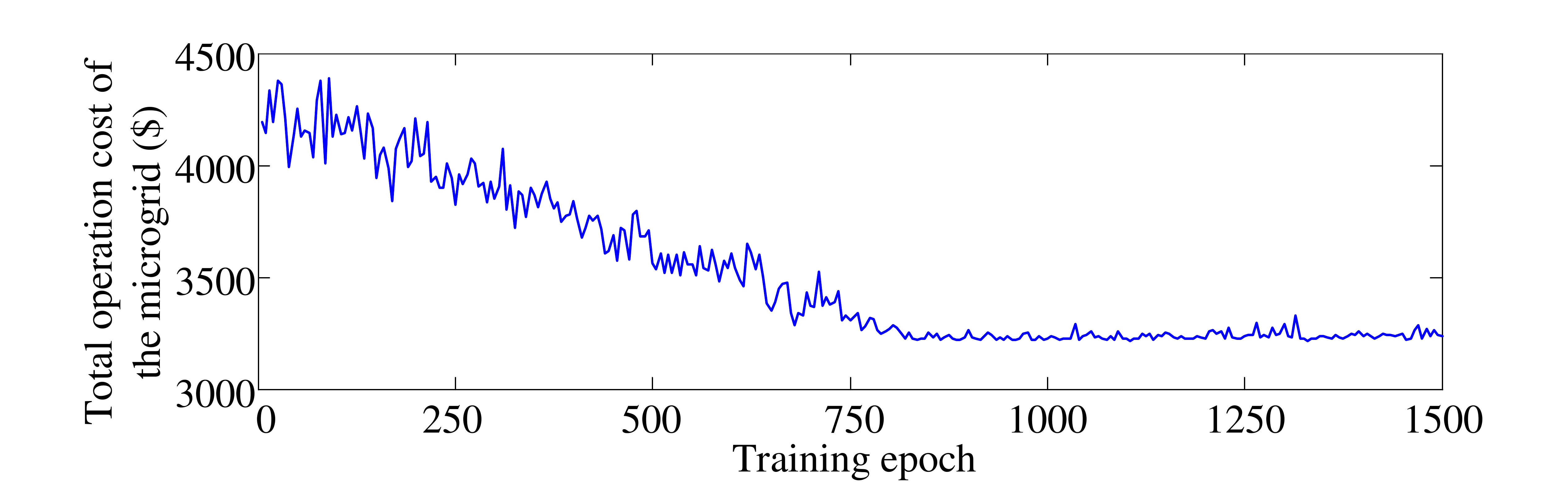}
\caption{The convergence process of the DDQN-RTO algorithm applied in the IEEE 69-bus microgrid test system.} \label{fig:IEEE69BusSystemCon}
\end{figure}

After 17 hours of training, the algorithm converged to a stable objective value.
The convergence curve of the proposed algorithm is shown in Fig. \ref{fig:IEEE69BusSystemCon}.
Using the agent trained off-line, we randomly selected 10 days which were not used in the training process to test the real-time optimization performance of the algorithm.
The results are shown in Table \ref{IEEE69}.
It can be found that the DDQN-RTO algorithm reaches a much smaller average optimality gap compared with the other two algorithms.
Besides, it takes 1.30 s to generate one schedule for the DDQN-RTO algorithm.
The time consumptions to generate one schedule for the ADP and the myopic policy are 13.46 s and 12.20 s, respectively.
One can find that the ADP and the myopic policy take a longer optimization time than the proposed algorithm.
This is because these two methods need to traverse all the feasible primary actions to select the optimal one, this process is time-consuming.
However, the proposed algorithm gets the optimal primary actions by Eq.(\ref{eq:primary actions making}) which corresponds to a neural network forward calculation procedure, then we solve an optimal power flow sub-problem to get the secondary actions.
Thus, the results validated the effectiveness of the proposed method on a relative larger test system.

\begin{table}
\newcommand{\tabincell}[2]{\begin{tabular}{@{}#1@{}}#2\end{tabular}}
\centering
\footnotesize
\caption{\label{IEEE69}The on-line optimization results of the DDQN-RTO, ADP, and Myopic policy on the modified IEEE-69 bus microgrid system}
\begin{tabular}{cccc}
\toprule
\tabincell{c}{Algorithm} &\tabincell{c}{DDQN-RTO} &\tabincell{c}{ADP} & \tabincell{c}{Myopic} \\ 	
\midrule
 \tabincell{c}{Average operation cost (\$)} & \tabincell{c}{\cellcolor[gray]{0.8}3196.3}	 & 3218.5 & 3248.3 \\
 \tabincell{c}{Average optimality gap (\%)} & \tabincell{c}{\cellcolor[gray]{0.8}4.33}	 & 5.06 & 6.03 \\
 \tabincell{c}{Maximum optimality gap (\%)} & \tabincell{c}{\cellcolor[gray]{0.8}7.86}	 & 8.11 & 10.26 \\
 \tabincell{c}{Minimum optimality gap (\%)} & \tabincell{c}{\cellcolor[gray]{0.8}3.19}	 & 3.27 & 4.25 \\
 \tabincell{c}{Standard deviation of \\ the optimization gaps ($10^{-2}$)} & $1.715$	 & \tabincell{c}{\cellcolor[gray]{0.8}$1.675$} & $2.092$ \\

\bottomrule
\end{tabular}
\end{table} 